\documentclass[aps,prb,reprint,superscriptaddress,showpacs,amsmath,amssymb,twocolumn]{revtex4-2}
\usepackage{graphicx}
\usepackage{dcolumn}
\usepackage{bm}
\usepackage{physics}
\usepackage{hyperref}
\usepackage{tikz}
\usepackage{booktabs}
\usepackage{multirow}

\definecolor{tudarkblue}{RGB}{0, 51, 102}    % Dark Blue for Nanowire
\definecolor{BLUE}{RGB}{0,0,250}
\begin{document}

\title{Disorder-robust trivial Majorana-like states from smooth confinement in chiral superconducting nanowires}
\author{Eslam Ahmed}
\email{ahmed.eslam@mbp.phys.kyushu-u.ac.jp}
\affiliation{Department of Physics, Kyushu University, Fukuoka, 819-0395, Japan}
\author{Jorge Cayao} 
 \email[ ]{ jorge.cayao@physics.uu.se}
\affiliation{Department of Physics and Astronomy, Uppsala University, Box 516, S-751 20 Uppsala, Sweden}
\author{Yukio Tanaka}
\email[]{tanaka.yukio.j2@f.mail.nagoya-u.ac.jp}
\affiliation{Department of Applied Physics, Nagoya University, Nagoya 464--8603, Japan}

\begin{abstract}
Near-zero-energy states in Majorana nanowires can arise from topologically trivial mechanisms such as smooth spatial inhomogeneity and disorder, making zero-energy pinning alone insufficient evidence of bulk topology. Here we identify a real-space mechanism governing their robustness to symmetry-preserving disorder. For a chiral-symmetric Bogoliubov–de Gennes Hamiltonian, we decompose a low-energy state into two normalized components of opposite chirality and show that disorder-induced splitting is bounded by their spatial overlap. We demonstrate this result in a finite Rashba nanowire with smooth chemical potential and pairing profiles. Below the bulk topological transition, smooth confinement produces partially separated chiral components with exponentially small overlap, yielding globally trivial Majorana-like Andreev bound states that remain near zero energy even under strong scalar, nonmagnetic disorder. The chiral overlap therefore provides a direct diagnostic of the protection of low-energy states against local perturbations, independent of the bulk topological invariant.

\end{abstract}

\maketitle
\section{Introduction}

Semiconductor--superconductor nanowires with Rashba spin-orbit coupling and a Zeeman field are among the most studied platforms for realizing Majorana bound states (MBSs) both theoretically \cite{Kitaev_2001,Lutchyn,Oreg,Alicea_2012,tanaka2012symmetry,Beenakker_2013,Stanescu_2013,sarma2015majorana,Aguadoreview17,sato2017topological,lutchyn2018majorana,prada2019andreev,flensberg2021engineered,Klinovaja_review2021,Marra_2022,tanaka2024theory} and experimentally \cite{zhang2019next,frolov2019quest,pita2025novel}.  In the idealized uniform limit, increasing the Zeeman field drives the proximitized wire through a bulk topological phase transition.  Above the transition, a pair of zero-energy MBSs appears at the two ends of the wire.  Because these states are spatially separated and protected by the superconducting gap, their energy splitting is exponentially small and insensitive to local perturbations that do not close the gap or break the protecting symmetries.

This expectation has made disorder robustness an important practical indicator in the search for Majorana zero modes \cite{ahmed2025anomalous}.  If a zero-energy state remains pinned near zero energy in the presence of local disorder, it is tempting to interpret this robustness as evidence for topological protection.  However, realistic nanowires are never perfectly uniform.  Smooth confinement \cite{PhysRevB.86.100503,marra2022majorana,prada2019andreev,PhysRevB.98.235406,PhysRevB.91.024514,PhysRevB.86.180503,baldo2023zero}, quantum dots \cite{PhysRevB.98.245407,PhysRevB.104.134507,PhysRevLett.123.117001}, nonuniform proximity coupling \cite{PhysRevB.105.144509}, interface profiles \cite{marra2022majorana,PhysRevB.102.245431,PhysRevB.104.L020501,prodanov2026interfaces}, non-Hermitian effects \cite{JorgeEPs}, and disorder \cite{Bagrets:PRL12,Pikulin2012A,DasSarma2021Disorder,PhysRevB.105.205122} can all generate low-energy Andreev bound states (ABSs) that mimic Majorana signatures.  In particular, partially separated ABSs can appear in locally topological regions produced by spatial inhomogeneity and can remain close to zero energy over extended parameter ranges \cite{PhysRevB.86.100503,prada2019andreev,PhysRevB.98.235406,marra2022majorana}.  These developments show that zero-energy pinning by itself is not a unique fingerprint of a globally topological phase.

Disorder itself has also played a central role in the Majorana-nanowire literature.  Scalar disorder, corresponding to nonmagnetic impurities or electrostatic potential fluctuations, is believed to be unavoidable in realistic superconductor--semiconductor devices and can destroy the energy gap, induce subgap states, and/or generate an accumulation of near-zero-energy states close to the topological transition  \cite{Bagrets:PRL12,PhysRevB.84.144526,PhysRevLett.106.057001,PhysRevB.83.184520,PhysRevB.85.140513,PhysRevLett.109.227006,Pikulin2012A,PhysRevB.88.064506,PhysRevB.94.140505,PhysRevB.107.184519,PhysRevB.105.205122}.  In most contexts, this effect is viewed as harmful because it obscures the distinction between true Majorana bound states and trivial low-energy states.  Since disorder and trivial low-energy ABSs are unavoidable in Majorana devices, it is natural to wonder under what conditions  such ABSs can remain robust against strong scalar disorder, and what microscopic mechanism is responsible for such a robustness?

In this work, we address these questions  from the viewpoint of chiral symmetry in Majorana devices where disorder, ABSs, and Majorana states coexist.  We show that, in a chiral-symmetric system, robustness against chiral symmetry-preserving disorder is controlled not only by the bulk topological invariant but also by the real-space overlap of the two opposite-chirality components of the low-energy wavefunction.  A chiral-symmetric local perturbation has no matrix elements within a fixed chirality sector.  It can split a near-zero mode only by coupling the two opposite chiralities.  Therefore, when these two chiral components are spatially separated, the disorder-induced splitting is suppressed even if the state is not protected by a global bulk topological phase.  In this sense, smooth confinement can create locally Majorana-like, but globally trivial, states whose disorder robustness is controlled by wavefunction properties rather than by a bulk invariant. This provides a natural way to understand how a trivial ABS can become robust.  The state may be topologically trivial in the global sense because the bulk has not undergone a topological phase transition.  At the same time, the spatial profile of the chemical potential and pairing can create a local Majorana-like wavefunction whose two chiral components are weakly overlapping.  Such a state is not a conventional topological MBS, but it can nonetheless be protected against chiral-symmetric local disorder by the same matrix-element suppression mechanism that protects spatially separated Majoranas.

\begin{figure}[t!]
    \centering
    \begin{center}
    \begin{tikzpicture}[scale=0.7]
        % Nanowire
        \fill[tudarkblue!20, rounded corners=3] (0,0) rectangle (10, 0.5);
        \draw[tudarkblue, thick, rounded corners=3] (0,0) rectangle (10, 0.5);
        \node at (5, 0.25) {Nanowire ($L$)};
        
        % Mu profile
        \draw[->] (0, 1) -- (10.5, 1) node[right] {$x$};
        \draw[->] (0, 0.8) -- (0, 2.2) node[left] {$\mu(x)$};
        \draw[thick, blue, domain=0:10] plot (\x, {1.3+1*(0.5 + 0.5*tanh((\x-5)/0.8))});
        \draw[<->, blue] (4.2, 1.75) -- (5.8, 1.75) node[midway, below] {$W_\mu$};
        
        % Delta profile
        \draw[->] (0, 2.7) -- (10.5, 2.7) node[right] {$x$};
        \draw[->] (0, 2.5) -- (0, 3.9) node[left] {$\Delta(x)$};
        \draw[thick, green!50!black, domain=0:10] plot (\x, {2.7+1*(0.5 + 0.5*tanh((\x-5)/1.5))});
        \draw[<->, green!50!black] (3.5, 3.25) -- (6.5, 3.25) node[midway, below] {$W_\Delta$};
    \end{tikzpicture}
    \end{center}
    \caption{Schematic sketch of the system.  A 1D Rashba nanowire of length $L$ with inhomogeneous chemical potential $\mu$ and induced pair potential $\Delta$. The spatial profile of $\mu$ and $\Delta$ are displayed.} %$\chi$ is a numerically-obtained fixed length we used to derive the disorder-robustness critical Zeeman field $B_{c1}$.}
    \label{fig1}
\end{figure}
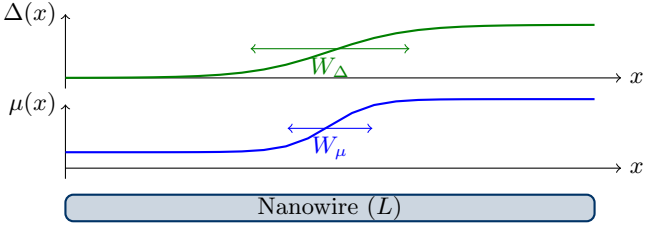

Thus, rather than taking disorder robustness as a signal of topology, we derive a simple bound showing when chiral-symmetric disorder is unable to split a low-energy state. To that end, we decompose the wavefunction into eigenstates of the chiral symmetry operator.  A similar decomposition based on the particle-hole symmetry operator was explored in previous studies on Majorana nanowires \cite{prada2019andreev,marra2022majorana,PhysRevB.98.235406,PhysRevB.110.165404,PhysRevLett.108.096802,PhysRevB.92.115115,PhysRevB.93.155425,bena_2017,Kaladzhyan_2017}. Our chiral symmetry-based decomposition is conceptually similar but is not directly related to the particle-hole Majorana decomposition used in the literature.  Our decomposition is designed to answer a different question: which part of the wavefunction can be coupled by a perturbation that preserves chiral symmetry?  This allows us to connect the numerical robustness of trivial zero modes directly to symmetry-based arguments for zero-energy edge states in superconductors with chiral symmetry \cite{tanaka2012symmetry, sato2017topological,sato2011topology,Mizushima_review2016,PhysRevB.95.214503,PhysRevB.97.174501}, anomalous proximity effects \cite{Proximityp, PhysRevB.91.174511,PhysRevB.72.140503,PhysRevB.94.054512,PhysRevB.102.140505,NagaeFlatband2025,odd1,Mizushima2023}, and  spectral bulk-boundary correspondence \cite{Spectralbulk,PhysRevB.100.174512,PhysRevB.104.165125}. As a concrete example, we study a finite Rashba nanowire in which both the chemical potential $\mu(x)$ and induced pair potential $\Delta(x)$ vary smoothly across the normal/superconductor interface, see Fig.~\ref{fig1}.  This system is useful because it interpolates between two well-known limits: a sharp NS junction, where trivial interface ABSs are generally fragile, and a uniform superconducting wire, where robust MBSs appear only after the bulk topological transition.  In the intermediate inhomogeneous regime, we find a third possibility.  A robust zero-energy state appears below the bulk topological transition.  Its onset occurs at a crossover field $B_{c1}$ where the two chiral components of the lowest-energy state separate in real space.  The true bulk topological transition occurs only at a higher field $B_{c2}$.  Thus, the interval $B_{c1}<B<B_{c2}$ realizes disorder robustness without global topology.

The main messages of this paper are therefore the following.  First, disorder robustness is not, by itself, a sufficient diagnostic of a global topological phase.  Second, in chiral-symmetric systems, the relevant microscopic quantity is the inter-chirality overlap $\Omega$, which bounds the perturbation-induced splitting.  Third, smooth inhomogeneous Rashba nanowires provide a concrete realization in which a topologically trivial but chiral-separated ABS is robust against strong scalar disorder.  Finally, disorder should not be viewed only as a destructive ingredient: in the present system it can also reshape and effectively move the low-energy chiral components, without coupling them and splitting them away from zero energy.

The rest of this paper is organized as follows. In Sec.~\ref{sec:1}, we derive the chiral-overlap criterion for local perturbations.  In Sec.~\ref{sec:2}, we introduce the inhomogeneous Rashba nanowire model and specify the scalar-disorder perturbation used in the numerical calculations. In Sec.~\ref{sec:3}, we review the basics of the inhomogeneous Rashba nanowire model and show how smooth confinement generates chiral-separated trivial zero modes. In Sec. \ref{sec:4}, we show how these modes respond to strong scalar disorder.  In Sec.~\ref{sec:5}, we analyze the local chiral density to visualize the near zero-energy state wavefunctions.  Finally, in Sec.~\ref{sec:6}, we summarize our work.

\section{Chiral-overlap criterion for disorder robustness}
\label{sec:1}
In this section, we establish a criterion for robustness against local chiral-symmetric perturbations. Since this criterion relies only on chiral symmetry and locality of perturbations, our criterion generalizes to systems beyond the specific semiconductor--superconductor heterostructure studied below and can also be used in non-superconducting chiral systems.  However, because our main application is to superconducting nanowires, we formulate the argument for an arbitrary BdG Hamiltonian $H$ with a unitary chiral symmetry operator $\Gamma$ satisfying
\begin{equation}
    \Gamma^2=1,\qquad \acomm{H}{\Gamma}=0.
    \label{eq:chiral_symmetry_general}
\end{equation}
Now, let us consider an arbitrary energy eigenstate of the Hamiltonian, $\psi_E$. Without loss of generality, we assume that $E$ is a positive eigenvalue.
\begin{equation}
    H\psi_E=E\psi_E,\qquad E>0.
\end{equation}
Chiral symmetry [Eq.\,\eqref{eq:chiral_symmetry_general}] implies that the state $\psi_E$ can never be common eigenstate of both $H$ and $\Gamma$ unless $E=0$ \cite{sato2011topology}. In particular, we have
\begin{equation}
    0=\expval{\acomm{H}{\Gamma}}{\psi_E}=2E\expval{\Gamma}{\psi_E},
\end{equation}
so that $\expval{\Gamma}{\psi_E}=0$ for non-zero energy eigenvalue $E$. Note that since $\Gamma^2 = 1$, the chiral operator has eigenvalues $\pm1$. To project the energy eigenstate $\psi_E$ into the two chiral sectors, we define the following projectors
\begin{equation}
    P_\pm=\frac{1\pm\Gamma}{2}.
\end{equation}
Since $\Gamma$ has zero expectation value for a given non-zero energy eigenstate $\psi_E$, it is straightforward to see that
\begin{equation}
    \norm{P_+\psi_E}^2=\norm{P_-\psi_E}^2=\frac{1}{2}.
\end{equation}
It is then natural to define normalized chiral components
\begin{equation}
    \phi_+=\sqrt{2}P_+\psi_E,
    \qquad
    \phi_- =\sqrt{2}P_-\psi_E,
    \label{eq:normalized_chiral_components}
\end{equation}
which satisfy
\begin{equation}
    \Gamma\phi_\pm=\pm\phi_\pm,
    \qquad
    \psi_E=\frac{1}{\sqrt{2}}\left(\phi_++\phi_-\right).
\end{equation}
The chiral symmetry forces $H$ to be off-diagonal in this basis.  Indeed,
\begin{equation}
    H\phi_+=E\phi_- ,
    \qquad
    H\phi_-=E\phi_+,
    \label{eq:H_maps_chiral_sectors}
\end{equation}
and therefore
\begin{equation}
    E=\left|\mel{\phi_+}{H}{\phi_-}\right|,
    \qquad
    \mel{\phi_\pm}{H}{\phi_\pm}=0.
    \label{eq:energy_chiral_matrix_element}
\end{equation}
Thus, the finite energy of the quasiparticle may be interpreted as the hybridization energy between two opposite-chirality components.

Now consider a perturbation $V$ that preserves the same chiral symmetry,
\begin{equation}
    \{V,\Gamma\}=0.
    \label{eq:V_chiral_symmetric}
\end{equation}
Such a perturbation is also off-diagonal in the chiral basis:
\begin{equation}
    \mel{\phi_\pm}{V}{\phi_\pm}=0.
\end{equation}
Consequently, it can shift the energy of the eigenstate only through
\begin{equation}
    M_V=\mel{\phi_+}{V}{\phi_-}.
    \label{eq:inter_chirality_matrix_element}
\end{equation}
For example, the first-order energy shift of $\psi_E$ is
\begin{equation}
    \delta E^{(1)}
    =\mel{\psi_E}{V}{\psi_E}
    =\Re\mel{\phi_+}{V}{\phi_-}.
\end{equation}
The same matrix element controls the perturbative splitting of the associated near-zero pair.  Therefore, the robustness of the state against a chiral-symmetric perturbation is controlled by the size of the matrix element $M_V$ that couples the two opposite chiralities.

For local perturbations, this matrix element is bounded by a simple real-space overlap.  We write the wavefunction at position $x$ as a superposition of two chiral components $\phi_\pm(x)$, and define the normalized local chiral densities
\begin{equation}
    \rho_\pm(x)=\phi_\pm^\dagger(x)\phi_\pm(x),
    \qquad
    \sum_x\rho_\pm(x)=1.
\end{equation}
Assume that $V$ is local,
\begin{equation}
    V=\sum_x V(x)\ketbra{x}{x},
\end{equation}
where $V(x)$ is a position-dependent matrix operator in the internal (particle-hole, spin, orbital, etc) degrees of freedom. 

We define the operator norm of the local blocks $V(x)$ as follows
\begin{equation}
    \norm{V(x)} = \max_{\lambda(x)\in {\rm eig}\left(V(x)\right)}\left|\left(\lambda(x)\right)\right|,
\end{equation}
where $\lambda(x)\in{\rm eig}\left(V(x)\right)$ is an eigenvalue of the matrix $V(x)$.

In addition to locality, we assume that the operator norms of all local blocks are bounded by
\begin{equation}
    \norm{V(x)}\le V_{\max}\qquad\text{for all } x.
\end{equation}
Then
\begin{align}
    \left|\mel{\phi_+}{V}{\phi_-}\right|
    &\le \sum_x \left|\phi_+^\dagger(x)V(x)\phi_-(x)\right| \\
    &\le \sum_x \norm{\phi_+(x)}\,\norm{V(x)}\,\norm{\phi_-(x)} \\
    &\le V_{\max}\sum_x\sqrt{\rho_+(x)\rho_-(x)}.
\end{align}
We therefore define the chiral overlap $\Omega$ as
\begin{equation}
    \Omega=\sum_x\sqrt{\rho_+(x)\rho_-(x)},
    \label{eq:Omega_definition}
\end{equation}
which gives the bound
\begin{equation}
    \left|\delta E^{(1)}\right|\leq\left|\mel{\phi_+}{V}{\phi_-}\right|
    \le V_{\max}\Omega .
    \label{eq:Omega_bound}
\end{equation}
The quantity $\Omega$ satisfies $0\le\Omega\le1$.  It is close to unity when the two chiral components occupy the same spatial region and close to zero when they are spatially separated.  Equation~\eqref{eq:Omega_bound} is the central criterion of this work: a state with small chiral overlap is insensitive to any local perturbation that preserves the chiral symmetry, even if the state is not protected by a global topological invariant.

It is useful to compare Eq.~\eqref{eq:Omega_definition} with overlap measures used previously for partially separated ABSs.  Earlier works often decompose a low-energy quasiparticle into two Majorana components related by particle-hole symmetry and quantify their spatial separation \cite{prada2019andreev,marra2022majorana}.  Our definition is closely related, but not equivalent to the one previously seen in the literature in general.  Instead, it is the chiral analogue appropriate for the present question.  The Majorana decomposition diagnoses the self-conjugate real-space structure of a quasiparticle.  The chiral decomposition diagnoses which part of the wavefunction can be coupled by a perturbation that anticommutes with $\Gamma$.  For this reason, $\Omega$ directly appears in the bound Eq.~\eqref{eq:Omega_bound}.  Whether the two overlap definitions become identical depends on additional properties of the low-energy subspace and on the representation of particle-hole and chiral symmetries.  We do not assume such an identity here; the protection mechanism follows directly from chiral symmetry and locality.

We emphasize that Eq.~\eqref{eq:Omega_bound} is not a statement only about scalar disorder.  It applies to any local perturbation whose BdG matrix anticommutes with the chiral operator.  Therefore, the same mechanism can protect a chiral-separated state against certain magnetic impurity configurations as well, provided that the impurity Hamiltonian preserves the same chiral symmetry.  Conversely, perturbations that break chiral symmetry are not constrained by this bound and can generally split the near-zero mode more efficiently, including topologically protected Majorana modes.  This distinction will be important when interpreting robustness in the presence of nonmagnetic versus magnetic disorder.

To end this section, we note that Symmetry-based arguments of this kind have played an important role in the theory of robust zero-energy states, anomalous proximity effects, and bulk-boundary correspondence in superconducting systems \cite{tanaka2012symmetry,Spectralbulk,PhysRevB.91.174511,Proximityp,PhysRevB.71.094513,odd1,odd3,PhysRevB.87.104513,PhysRevB.94.054512,PhysRevB.95.214503,PhysRevB.97.174501,PhysRevB.102.140505,NagaeFlatband2025,kokkeler2022,tanaka2024theory}. 
\section{Inhomogeneous Rashba nanowire model}
\label{sec:2}
We now demonstrate the criterion in Eq.\,\eqref{eq:Omega_bound} using a finite Rashba nanowire with spatially inhomogeneous chemical potential and induced superconducting pair potential.  We model Majorana nanowire devices under realistic experimental conditions. To that end, we consider a finite one-dimensional Rashba nanowire of length $L$ with strong Rashba spin-orbit coupling. The right half of the wire is brought into proximity to a conventional $s$-wave superconductor which induces Cooper pairing correlations inside the right half of the wire via the proximity effect. Moreover, an external Zeeman field perpendicular to the spin-orbit coupling axis is applied which can induce a topological phase transition inside the wire. To model realistic experimental scenarios, we assume that the wire is connected to voltage  gates which affect the magnitude of the chemical potential inside the wire. We also assume that the effect of the voltage gates is weak inside the superconductor, thus, creating an inhomogeneous chemical potential profile inside the wire. This implies that the chemical potential smoothly changes from one value deep in the superconducting region to a different much lower value deep in the normal region. Moreover, we assume the same inhomogeneity in the value of the pair potential inside the wire such that the pair potential doesn't suddenly drop to zero at the normal/superconductor (NS) interface. As a result, both the effective chemical potential $\mu(x)$ and the induced pair potential $\Delta(x)$ vary smoothly over interface widths $W_\mu$ and $W_\Delta$, respectively as seen in Fig.~\ref{fig1}.. 

Overall, the wire is modeled by the following Bogoliubov-de Gennes (BdG) Hamiltonian:
\begin{equation} 
    H = \left(\frac{p^2}{2m} - \mu(x)\right)\tau_z + \alpha p_x \sigma_z + B\sigma_x\tau_z + \Delta(x)\sigma_y\tau_y ,
    \label{eq:BdG_cont}
\end{equation}
where $\sigma_i$ ($\tau_i$) are Pauli matrices in spin (Nambu) space. The Hamiltonian has chiral symmetry
\begin{equation}
    \{H,\Gamma\}=0,
    \qquad
    \Gamma=\sigma_x\tau_y.
    \label{eq:nanowire_chiral_symmetry}
\end{equation}
This symmetry places the clean model in class BDI \cite{tanaka2012symmetry}. Here, both the chemical potential and pair potential have inhomogeneity which we model using a smooth step function. In particular, we consider the following function:
\begin{equation}
    u(x) = \frac{1}{2}\left(  1 + \tanh(x) \right).
\end{equation}
Using this function, we define the chemical potential and pair potential as follows:
\begin{align}
    \mu(x) &= \mu_L + \left(\mu_R-\mu_L\right)u\left(\frac{x-x_0}{W_\mu}\right),
    \\
    \Delta(x) &= \Delta_0u\left(\frac{x-x_0}{W_\Delta}\right)
\end{align}
where $x_0=L/2$ denotes the NS interface position. We have $\mu_{R/L}=\mu(x=\pm\infty)$ is the values of the chemical potential far to the left or far to the right and $\Delta_0=\Delta(\infty)$ is the value of the pair potential at the far right. Moreover, $W_\mu$ and $W_\Delta$ control the width of the region where $\mu$ and $\Delta$ significantly change value, respectively. Note that since the wire has finite length, the actual values of the chemical potential and pair potential are bounded to $\mu\in\left[\mu(0),\mu(L)\right]$ and $\Delta\in\left[\Delta(0),\Delta(L)\right]$, respectively. The limits $W_\mu=W_\Delta=0$ and $W_\mu=W_\Delta\to\infty$ correspond to a NS junction with sharp interface and a uniform superconducting wire, respectively (Fig.~\ref{fig2}). Note that in the uniform superconducting wire limit, the values of the chemical potential and pair potential are equal to the average values ($\mu_{uniform} = (\mu_R+\mu_L)/2, \Delta_{uniform}= \Delta_0/2$). 

To move forward, we discretize Eq.~(\ref{eq:BdG_cont}) on a 1d lattice with lattice spacing $a$ and length $N=L/a$ lattice points. The kinetic term is implemented with nearest-neighbor hopping $t=\hbar^2/(2ma^2)$, and the Rashba term is discretized by a symmetric finite difference. In Nambu basis \(\Psi_j=(c_{j\uparrow},c_{j\downarrow},c^\dagger_{j\uparrow},c^\dagger_{j\downarrow})^T\), the tight-binding Hamiltonian reads
\begin{align}
H &= \frac{1}{2}\sum_{j}\Psi_j^\dagger\Big[(2t-\mu_j)\tau_z + B\,\sigma_x\tau_z + \Delta_j \sigma_y\tau_y \Big]\Psi_j \nonumber\\
&\quad + \frac{1}{2}\sum_{j}\Big[\Psi_{j+1}^\dagger\Big(-t\,\tau_z -\frac{i\alpha}{2a}\sigma_z\Big)\Psi_j + \mathrm{H.c.}\Big].
\end{align}
To probe stability against imperfections, we add on-site scalar disorder to the left side of the wire:
\begin{equation}\label{eq:H_dis}
H_{\mathrm{dis}}=\frac{1}{2}\sum_j \Psi_j^\dagger \, V_j \tau_z \, \Psi_j,
\end{equation}
where \(V_j\) is drawn independently from a uniform distribution \(V_j\in[-V_0,V_0]\).  This term represents scalar, nonmagnetic disorder, or equivalently local electrostatic potential fluctuations.  Such disorder is especially relevant for superconductor--semiconductor nanowires because it is difficult to eliminate experimentally and is known to generate subgap states and zero-bias conductance peaks in realistic devices \cite{PhysRevB.107.184519,PhysRevB.94.140505,PhysRevB.105.205122,DasSarma2021Disorder,Bagrets:PRL12,PhysRevB.84.144526,PhysRevLett.106.057001,PhysRevB.83.184520,PhysRevB.85.140513,PhysRevLett.109.227006,Pikulin2012A,PhysRevB.88.064506,PhysRevB.94.140505,PhysRevB.107.184519,PhysRevB.105.205122}. Importantly, the scalar disorder in Eq.~\eqref{eq:H_dis} preserves the chiral symmetry because
\begin{equation}
    \{\tau_z,\Gamma\}=0.
\end{equation}
Thus,  disorder can split a low-energy state only through the inter-chirality matrix element bounded in Eq.~\eqref{eq:Omega_bound}.  In the calculations below, the disorder is applied to the left half of the wire.  We consider two limits: the clean limit with $V_0=0$ and an extreme disorder limit with $V_0=50\mu_R$. We use this deliberately strong disorder strength because only states with a genuine symmetry-based suppression of the disorder matrix element remain near zero energy in this regime.  In the extreme disorder limit, we average over 10 disorder realizations. We have confirmed that increasing disorder realizations does not affect our results.

Although our numerical calculations focus on the scalar disorder, the chiral-overlap criterion in Eq.~\eqref{eq:Omega_bound} is more general.  Any local perturbation whose BdG matrix anticommutes with $\Gamma$ is constrained by the same bound.  For example, in the convention of Eq.~\eqref{eq:BdG_cont}, a fixed-direction magnetic impurity parallel to the Zeeman axis and represented by $b_x(x)\sigma_x\tau_z$ also satisfies $\{\sigma_x\tau_z,\Gamma\}=0$.  This magnetic perturbation is therefore chiral-symmetric and has to satisfy Eq.~\eqref{eq:Omega_bound}. Thus, any energy splitting of an energy eigenstate is bounded by the amount of overlap $\Omega$ between the chiral components of that eigenstate.  By contrast, generic magnetic disorder does not need to preserve chiral symmetry. In particular, components that do not anticommute with $\Gamma$ are outside the protection mechanism discussed here and can lift the near-zero mode more efficiently.

To end this section, we list the parameter values used in the numerical simulations presented below. We use realistic parameters consistent with experimental values for InSb and InAs nanowires \cite{lutchyn2018majorana}. To speed up numerical simulations, we consider a large lattice constant ($a=50$nm),  yielding a hopping parameter of $t=1$ meV. This was proved useful when studying other  properties of Rashba nanowire systems \cite{ahmed2025anomalous}. The rest of the  parameters are chosen as follows: the chemical potential on the far left $\mu_L=0.1$ meV, the chemical potential on the far right $\mu_R=0.5$ meV, the pair potential on the far right $\Delta_0=0.5$ meV, and Rashba spin-orbit coupling strength $\alpha=20$ meV nm. The total number of sites is $N=200$ sites which yields a total length of $L=10$ microns. Furthermore, we define $B_c=\sqrt{\mu_R^2 + \Delta_0^2}\approx 0.7$ meV which is the critical Zeeman field for the bulk topological phase transition in an infinite wire. Note that since our wire is finite and inhomogeneous, the actual critical Zeeman field which corresponds to the bulk gap closing point does not have to coincide with the infinite wire critical Zeeman field. Nevertheless, we can estimate the value of the Zeeman field at which the bulk gap closing occur. To that end, we define the value $B_{c2}=\sqrt{\mu(x=L)^2 +\Delta(x=L)^2}$ which is the critical Zeeman field above which the entire wire is guaranteed to be topological. We find that this value accurately predicts the actual bulk gap closing point. We define a lower field $B_{c1}$ which denotes the crossover at which the lowest-energy wavefunction becomes fully chiral-separated.  The central result is that disorder robustness begins at $B_{c1}$ rather than at $B_{c2}$.

\section{Rashba nanowire in the ballistic regime}\label{sec:3}
The ballistic limit of Rashba nanowires --including the case with inhomogeneous smoothly varying chemical potential and pairing potential-- have been extensively studied in the literature. For the sake of completeness, we review the basics of the ballistic limit in this section in light of chiral symmetry.

\subsection{Uniform S and NS junction limits}
Having introduced the theoretical model, we now analyze the energy spectrum of our model under different experimental constraints. For the sake of completeness, we first consider two limiting cases: $W_\mu=W_\Delta=\infty$ which corresponds to a uniform superconductor, and $W_\mu=W_\Delta=0$ which corresponds to an NS junction. These two limiting cases are well-understood in the literature with many works exploring them \cite{Oreg,Lutchyn,Alicea_2012,Beenakker_2013,sarma2015majorana,Aguadoreview17,sato2017topological,lutchyn2018majorana,flensberg2021engineered,tanaka2024theory,cayao2016hybrid,Marra_2022,PhysRevB.87.104513}. Below, we review these cases.

\begin{figure}[t!]
    \centering
    \includegraphics[width=1.0\linewidth]{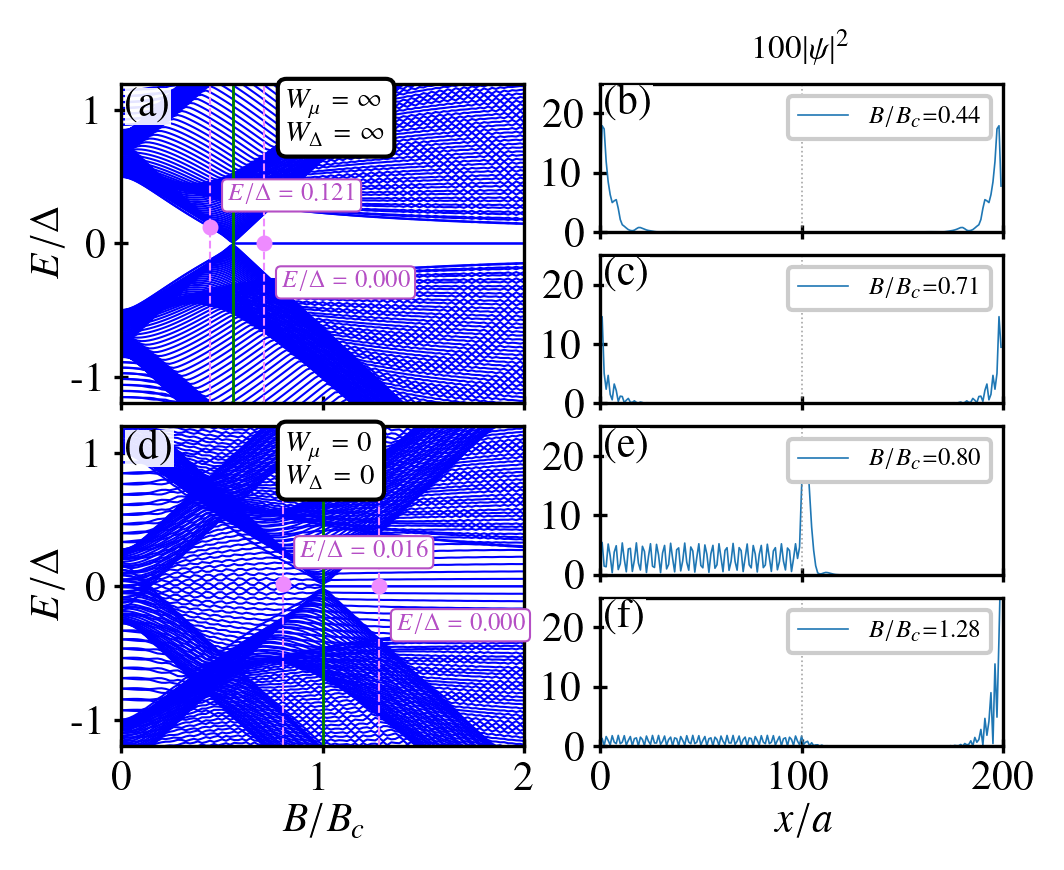}
    \caption{Low-energy spectrum for uniform superconductor with $W_\mu =W_\Delta=\infty$ (a) and for NS junction with $W_\mu =W_\Delta=0$ (d). The right panels show the lowest positive energy level wavefunction probability density at two fixed Zeeman fields for the uniform superconductor case (b,c) and NS junction case (e,f). The fixed Zeeman fields used to plot the wavefunctions are indicated by the pink dashed lines in panels (a, d). We also show the energies of the plotted wavefunctions. The green line indicates the bulk gap closing point $B_{c2}$. }
    \label{fig2}
\end{figure}

Fig.~\ref{fig2}(a) shows the low-energy spectrum for the uniform superconducting limit $(W_\mu=W_\Delta=\infty)$. In this limit, we see that $u(x)\to1/2$, implying $\mu(x)=(\mu_L+\mu_R)/2$ and $\Delta(x)=\Delta_0/2$ regardless of position $x$. 
As expected, the bulk gap closes at the bulk topological transition $B_{c2}=\sqrt{(\mu_L+\mu_R)^2+\Delta_0^2}\Big/2$ , indicated by the green line, after which a pair of end-localized Majorana bound states emerges and remains pinned exponentially close to zero energy in the finite system. We indeed confirm the existence of these Majorana bound states by checking their wavefunction probability density in Fig.~\ref{fig2} (c).

Below the critical Zeeman field $B_{c2}$, we have a completely gapped trivial superconductor, with a continuum of states above the energy gap as seen in Fig.~\ref{fig2} (a). We note that due to open-boundary condition at both ends, two pairs of degenerate finite-energy edge modes appear in the trivial regime, with one pair appearing at positive energy and another pair at negative energy. These two pairs are located near the energy continuum and are not possible to see in the spectrum in Fig.~\ref{fig2} (a). However, we can clearly see these finite energy edge modes by checking their probability densities in Fig.~\ref{fig2} (b). Indeed, we see that the probability densities of these states are localized at the edge of the superconductor, similar to that of the MBSs. However, unlike MBSs, these states have energy comparable to the bulk gap. We note that these states have been observed before in literature, see for example \cite{cayao2018andreev,ahmed2025odd} .

Next, we shift our attention to the NS regime $(W_\mu=W_\Delta=0)$. In this regime, only the right half of the wire is superconducting with $\mu(x>\frac{L}{2})=\mu_R$ and $\Delta(x>\frac{L}{2})=\Delta_0$. while the normal side on the left half of the wire has  $\mu(x<\frac{L}{2})=\mu_L$ and $\Delta(x<\frac{L}{2})=0$. We have addressed this scenario in previous work (see Refs.~\cite{ahmed2025odd, ahmed2025anomalous, cayao2016hybrid}) where we have found that due to finite size and the non-uniform nature of the chemical potential, both trivial zero energy Andreev bound states and topological Majorana bound states can appear in the same system under different parameters regimes, see Fig.~\ref{fig2} (d,e,f). In particular, we observe two bulk-gap closing points in the spectrum in Fig.~\ref{fig2} (d). The first bulk gap closing happens when the Zeeman field is equal to the chemical potential of the normal region, $B=\mu_L$. For $B>\mu_L$, the normal region becomes helical with only one active spin degree of freedom at low energies. In a previous work, we showed that in the helical phase, ABSs appear at the interface of the NS junction, see Fig.~\ref{fig2} (e). The energy levels of the ABSs oscillate as a function of Zeeman field and can mimic true MBSs in short nanowires \cite{ahmed2025odd}, however, we have shown that these states are fragile against even the smallest amount of disorder \cite{ahmed2025anomalous}. Another important feature of these trivial ABSs is their spatial localization as seen in Fig.~\ref{fig2} (e). In particular, the probability density shows a localized Gaussian-shaped spatial localization at the NS interface and an oscillatory behavior inside the N region. The second gap closing point happens at the topological phase transition of the superconductor at critical Zeeman field $B_{c2}=\sqrt{\mu_R^2+\Delta_0^2}$. For Zeeman fields above this critical value, a pair of MBSs appear at the edges of the superconducting region \cite{Lutchyn,Oreg}. We note that unlike the uniform superconductor case, the left MBS in the NS geometry leaks into the normal region due to the absence of a pairing gap $\Delta$ there, see Fig.~\ref{fig2}(f).

\subsection{ Rashba nanowires with smoothly varying chemical and pairing potentials}
\begin{figure}[t!]
    \centering
    \includegraphics[width=1.0\linewidth]{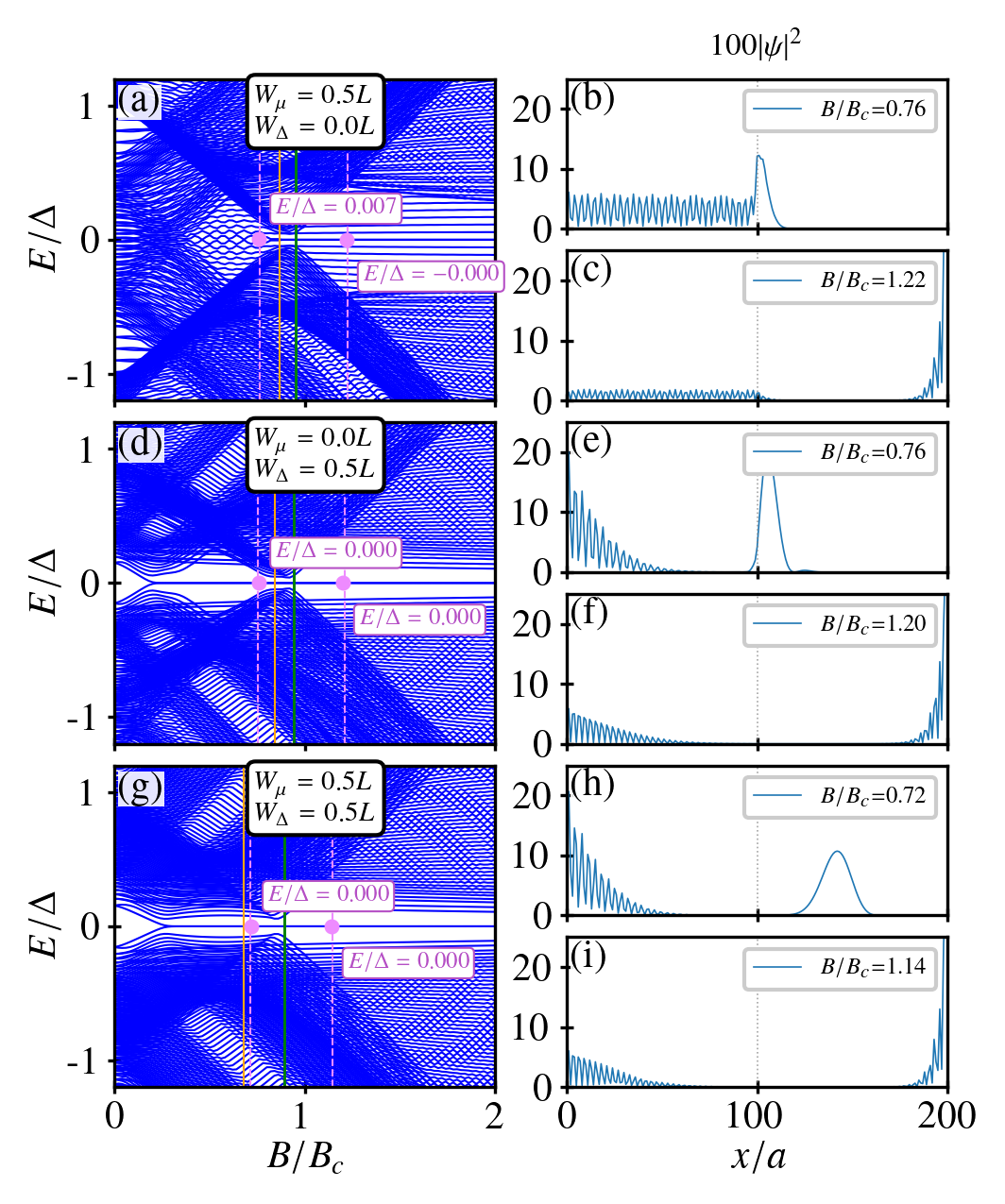}
    \caption{Low-energy spectrum for inhomogeneous nanowire with $W_\mu = 0.5L, W_\Delta=0$ (a), $W_\mu = 0, W_\Delta=0.5L$ (c), and $W_\mu =W_\Delta=0.5L$ (e). The right panels show the lowest positive energy level wavefunction probability density at two fixed Zeeman fields for $W_\mu = 0.5L, W_\Delta=0$ case (b),   $W_\mu = 0, W_\Delta=0.5L$ case (d), and $W_\mu =W_\Delta=0.5L$ case (f). The fixed Zeeman fields used to plot the wavefunctions are indicated by the pink dashed lines in panels (a, c). The green line indicates the bulk gap closing point, $B_{c2}$, while the orange line indicates the critical Zeeman field separating disorder-fragile and disorder-robust zero-energy states, $B_{c1}$.}
    \label{fig3}
\end{figure}

Having introduced the uniform superconductor and NS junction limits, we  next interpolate between these limits by introducing smooth profiles for the chemical potential and/or the induced pairing (Fig.~\ref{fig3}). When only $\mu(x)$ is smooth as in Fig.~\ref{fig3} (a,b,c) with $W_\mu=L/2, W_\Delta=0$, we see that the spectrum is roughly similar to that of the NS junction geometry in Fig.~\ref{fig2} (d) with only two differences. The first difference is that the topological gap closing happens at a much lower Zeeman field $B_{c2}$ than that in the NS junction case $(B_{c2}<B_c)$. The second difference is that the zero energy oscillations of the trivial ABS energy levels completely disappear above the orange line in Fig.~\ref{fig3} (a), way before the bulk gap closes.

\begin{figure*}[th!]
    \centering
    \includegraphics[width=1.0\linewidth]{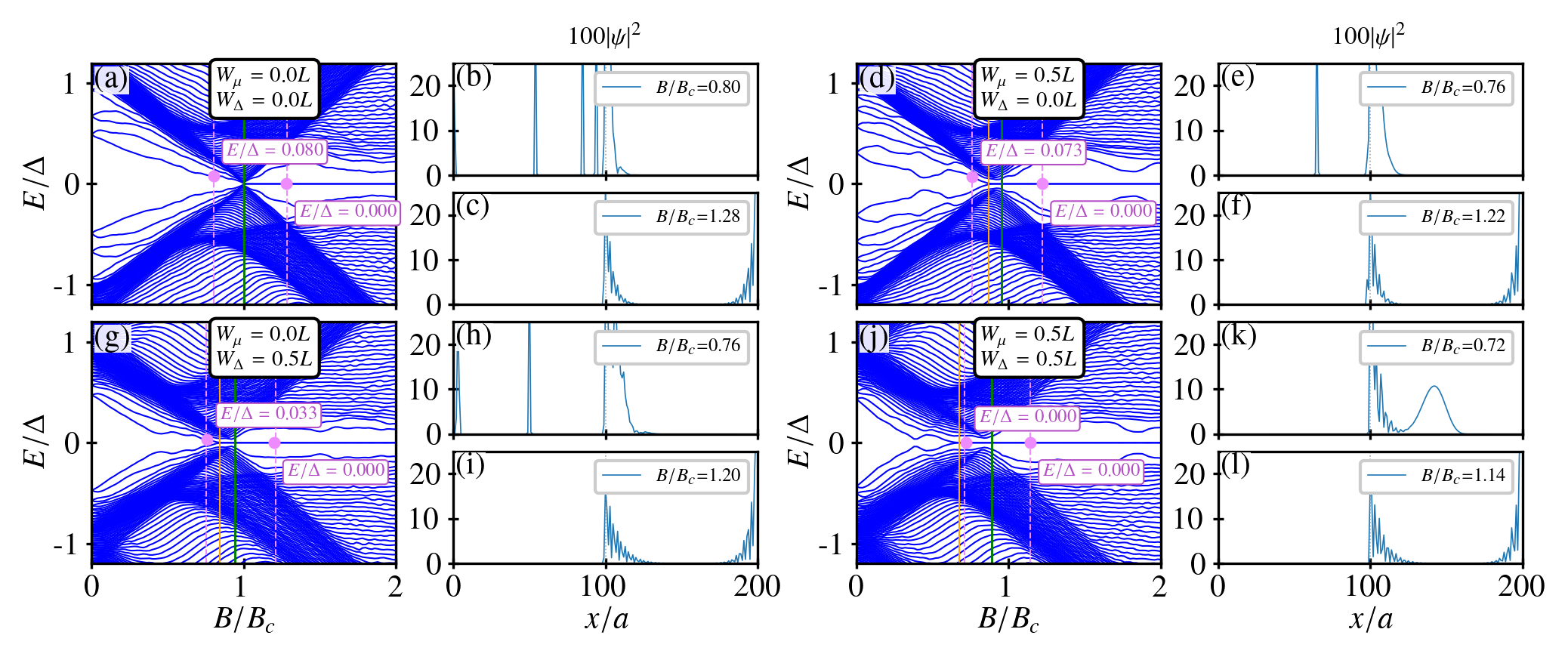}
    \caption{Impurity-averaged Low-energy spectrum for inhomogeneous nanowire in the presence of strong on-site disorder on the left half of the system with $W_\mu=W_\Delta=0$ (a), $W_\mu = 0.5L, W_\Delta=0$ (c), $W_\mu = 0, W_\Delta=0.5L$ (e), and $W_\mu =W_\Delta=0.5L$ (g). The remaining panels show the lowest positive energy level wavefunction probability density at two fixed Zeeman fields for $W_\mu=W_\Delta=0$ case (b), $W_\mu = 0.5L, W_\Delta=0$ case (d),   $W_\mu = 0, W_\Delta=0.5L$ case (f), and $W_\mu =W_\Delta=0.5L$ case (h). The fixed Zeeman fields used to plot the wavefunctions are indicated by the pink dashed lines in panels (a, c, e, g). The green line indicates the bulk gap closing point while the orange line indicates the critical Zeeman field separating disorder-fragile and disorder-robust zero-energy states.}
    \label{fig4}
\end{figure*}
Next, we consider the case with only $\Delta(x)$ being smooth as in Fig.~\ref{fig3} (d,e,f) with $W_\mu=0, W_\Delta=L/2$. In this case, we observe complete suppression of the trivial ABS zero-energy oscillations, see Fig.~\ref{fig3} (d). Instead, we have clear zero-energy trivial ABSs over a wide range of Zeeman fields. Moreover, the corresponding probability density of the trivial ABS ( Fig.~\ref{fig3} (e) ) becomes partially separated with one component shifting toward the left edge of the wire where the superconducting gap $\Delta(x)$ is minimum, and the other component remaining at the interface. The same effect is also observed in the topological phase where we observe that the left MBS is shifted towards the left end of the nanowire in an attempt to escape to a region with smaller superconducting gap $\Delta$, see Fig.~\ref{fig3} (f). On the other hand, the right MBS remains unchanged from the uniform superconductor and NS junction cases.

Similar results can be seen when both $\mu(x)$ and $\Delta(x)$ are smooth as in Fig.~\ref{fig3} (g,h,i) with the only difference being that the right component of the trivial zero-energy ABS is not located at the interface, but rather in the middle of the right side of the system. As we discussed in Sec. \ref{sec:1}, this enhanced separation between the two components will play an important role when we consider the effect of disorder. Importantly, we observe a sharp crossover at a Zeeman field $B_{c1}$ (orange line) below which the zero-energy mode is interface-like with significant spatial overlap. Meanwhile, above $B_{c1}$ the zero-energy mode becomes effectively split into two weakly overlapping components, suggesting enhanced robustness. Our results are consistent with previous studies in systems with smooth inhomogeneity \cite{PhysRevB.86.100503, PhysRevB.98.235406, marra2022majorana}.  
\section{Disorder effects on Rashba nanowires with smoothly varying chemical and pair potentials}\label{sec:4}
In the previous section, we have seen that smooth inhomogeneity in the chemical potential and pair potential leads to zero-energy trivial ABSs which mimics many signatures of truly topological MBSs. In our previous work, we showed that in NS junctions with a sharp step-function profile even small amount of disorder is sufficient to lift these trivial ABSs from zero energy \cite{ahmed2025anomalous}. Naively, one might therefore expect that the zero-energy ABSs in nanowires with smooth step-function profiles should also be fragile against disorder. However, as we discussed in Sec.\ref{sec:1}, any disorder-induced energy splitting should be suppressed if the chiral overlap of the wavefunction is very small. To test this hypothesis, we consider on-site scalar disorder in the left half of the nanowire (Fig.~\ref{fig4}). To insure that a gap opens for fragile states, we consider extremely strong disorder strength $V_0=50\mu_R$. 

Under this strong disorder, only truly disorder-robust states can survive. Indeed, we see that in the NS junction with sharp interface geometry [Fig.~\ref{fig4}(a,b,c)], disorder strongly perturbs the low-energy spectrum and readily shifts the near-zero-energy trivial ABSs away from zero energy. Meanwhile, topological MBSs are extremely robust and remain at zero energy. Disorder also has a strong impact on the spatial profile of both the trivial ABSs and topological MBSs. Due to strong disorder, the left half of the wire is effectively cut from the system and only the right side remains active in the low-energy regime. This is directly observed in the spatial distribution of the probability densities of both the trivial ABSs and topological MBSs, see Fig.~\ref{fig4} (b,c). In the trivial phase, the spatial oscillations of the probability density in the normal region is completely absent. The only feature that still remains is the Gaussian-shaped spatial distribution at the NS interface. In the topological phase, the right MBS remains unchanged while the left MBS cannot penetrate the N region anymore and is now localized at the interface and decays into the middle of the S region. Similarly, when only one of the profiles is smooth [Fig.~\ref{fig4}(d--i)], the near-zero-energy features remain fragile over much of the parameter range.

Strikingly, when $\mu(x)$ and $\Delta(x)$ are simultaneously smooth [Fig.~\ref{fig4}(j,k,l)], a wide window of Zeeman fields $B$ exists in which the lowest energy modes remain pinned exponentially close to zero energy even under strong disorder. Perhaps more surprisingly, the boundary separating the disorder-fragile and disorder-robust regimes does not coincide with any bulk-gap closing in the clean limit [Fig.~\ref{fig3} (g)]. Instead, this boundary coincides with the orange line $B_{c1}$, suggesting that the robustness is controlled not by the global topological transition but by the shape of the wavefunction, which we quantified in Sec. \ref{sec:1} using chiral symmetry. To see that, it is enlightening to consider the spatial distribution of the wavefunction in this topologically-trivial disorder-robust regime. In Fig.~\ref{fig4} (k), we can still identify two spatially-separated wavefunction components. Interestingly, the left component resembles that of a Majorana mode localized at the system's end. On the other hand, the right component is Gaussian-shaped. This partial separation between the right and left components further cements the chiral symmetry-based diagnostics we established in Sec. \ref{sec:1}.

\begin{figure}[b!]
    \centering
    \includegraphics[width=1.0\linewidth]{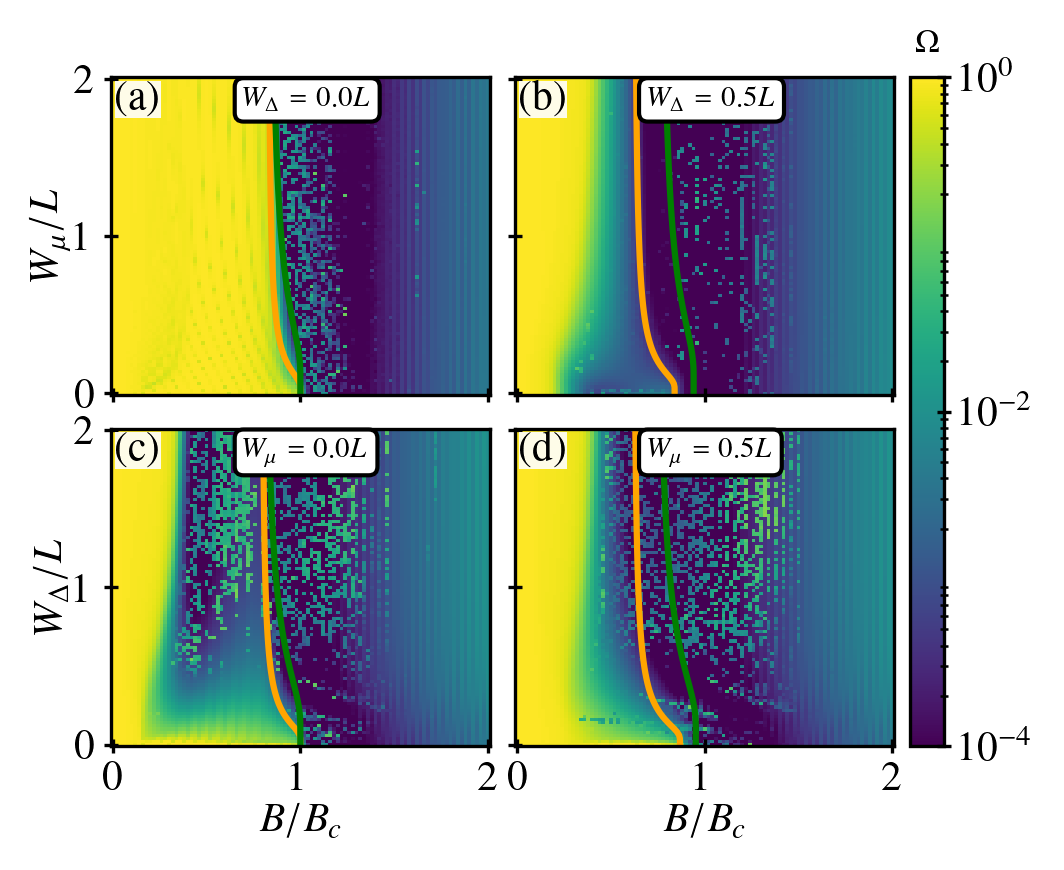}
    \caption{Spatial overlap ($\Omega$) between the two chiral components of the lowest positive energy level as a function of applied Zeeman field and $W_\mu$ for $W_\Delta=0$ (a), and for $W_\Delta=0.5L$ (b) and as a function of applied Zeeman field and $W_\Delta$ for $W_\mu = 0$ (c) and $W_\mu=0.5L$ (d). The green line indicates the bulk gap closing point while the orange line indicates the critical Zeeman field separating disorder-fragile and disorder-robust zero-energy states. Note that the value of the lowest positive energy level is not constant and depends on $B, W_\mu,$ and $W_\Delta$.}
    \label{fig5}
\end{figure}

To fully support our hypothesis, we now show that the chiral overlap $\Omega$ in Eq.~\eqref{eq:Omega_definition} fully characterizes disorder-robustness and provides a direct microscopic interpretation of the critical field $B_{c1}$ introduced earlier from the energy spectrum data. To see that, we begin with Fig.~\ref{fig5}, which shows the overlap $\Omega$ in the clean system as a function of Zeeman field and inhomogeneity width. First, we focus on the case where the pairing profile is sharp, $W_\Delta=0$, while the smoothness of the chemical potential is varied which is shown in Fig.~\ref{fig5} (a). As expected, for large Zeeman fields in the topological regime, the overlap is small as a result of Majorana non-locality. On the other hand, we see that at Zeeman fields smaller than the topological bulk gap closing field $B<B_{c2}$, the chiral components of the lowest energy state have huge overlap $\Omega\approx1$. At Zeeman fields close to $B_{c2}$ (the green line) and intermediate values of $W_\mu$, there is a very small region where the overlap drop significantly. This region bounded between the orange line $B_{c1}$ and the green line $B_{c2}$ marks the disorder-robust trivial regime. This demonstrates that a smooth chemical potential profile by itself already promotes some degree of chiral separation. Nevertheless, the low-overlap region remains relatively narrow, indicating that smooth $\mu(x)$ alone is not sufficient to produce disorder-robust trivial states.

Fig.~\ref{fig5}(c) shows the complementary situation in which the chemical potential remains sharp while the pairing profile is varied. Interestingly, the overlap decreases with increasing Zeeman field and increasing $W_\Delta$ even when the Zeeman field is far smaller than the bulk gap closing critical field $B_{c2}$. Physically, this happens because a finite $W_\Delta$ implies that the entire system is superconducting as there is now a finite non-zero $\Delta(x)>0$ for all $x$. In this situation, one chiral component drifts toward the side of the wire with smaller energy gap $\Delta$, thereby reducing its overlap with the other component which remains at the NS interface.  Compared with Fig.~\ref{fig5}(a), the suppression of overlap is stronger. We note that this suppression of overlap in the trivial regime seen in Fig.\,\ref{fig5}(c) can be explained from Fig.~\ref{fig3}(e). There, we clearly see the wavefunction is completely separated into two components in real space.
Even so, this overlap suppression remains incomplete over a substantial part of the trivial regime.

\begin{figure}[t!]
    \centering
    \includegraphics[width=1.0\linewidth]{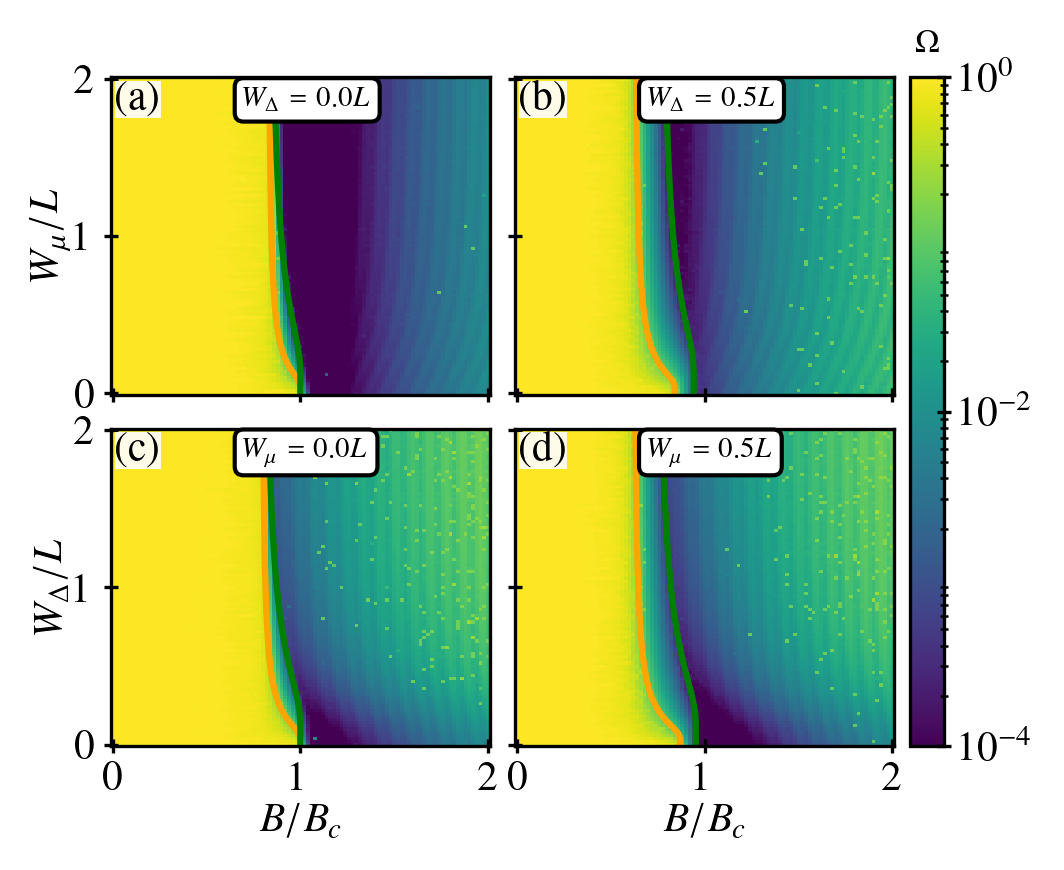}
    \caption{Impurity-averaged spatial overlap ($\Omega$) between the two chiral components of the lowest positive energy level in the presence of strong disorder as a function of applied Zeeman field and $W_\mu$ for $W_\Delta=0$ (a), and for $W_\Delta=0.5L$ (b) and as a function of applied Zeeman field and $W_\Delta$ for $W_\mu = 0$ (c) and $W_\mu=0.5L$ (d). The green line indicates the bulk gap closing point while the orange line indicates the critical Zeeman field separating disorder-fragile and disorder-robust zero-energy states. Note that the value of the lowest positive energy level is not constant and depends on $B, W_\mu,$ and $W_\Delta$.}
    \label{fig6}
\end{figure}

The most significant behavior of the spatial overlap quantifier $\Omega$ appears in Figs.~\ref{fig5}(b) and \ref{fig5}(d), where one parameter is already smooth and the smoothness of the other parameter is varied. In both cases, once the second parameter also becomes smooth, a broad dark region with very small overlap develops well below the green line, marking the bulk topological transition. The orange line tracks the left boundary of this low-overlap region, showing that $B_{c1}$ is naturally identified as the onset of chiral separation. In other words, the field $B_{c1}$ marks the point where the trivial low-energy mode stops behaving like a conventional interface ABS and starts behaving like a fully separated Majorana-like state. This interpretation immediately explains why the orange line in the energy spectrum figures does not coincide with the bulk-gap closing. The quantity that changes at $B_{c1}$ is not the bulk topology of the entire wire, but the internal geometry of the low-energy wavefunction. Below $B_{c1}$, the two chiral sectors occupy essentially the same region of space, so any perturbation that couples them can easily shift the mode away from zero energy. Above $B_{c1}$, the two sectors become weakly overlapping, and their hybridization is therefore strongly suppressed. This is precisely the regime where disorder robustness becomes possible even without a global topological transition. We note that this is similar to what happens on the dirty surface of nodal superconductors \cite{PhysRevB.90.064513,PhysRevB.95.214503}.

Further insights on the response of $\Omega$ against strong disorder is obtained from Fig.~\ref{fig6}, which shows the same overlap of the chiral components of the lowest positive energy state in the presence of very strong disorder on the left half of the wire. The comparison with Fig.~\ref{fig5} is highly revealing. Our first observation is that strong disorder increases the overlap between the chiral components across the entire parameters regimes. This is a direct result of the fact that the left half of the wire became effectively an insulator and thus inaccessible to the wavefunction which forces the chiral components to get closer to each other. Nonetheless, the overlap still remains small for Zeeman fields above the orange line ($B>B_{c1}$). Figs.~\ref{fig6}(b,d) provide the clearest evidence for our interpretation. When both $\mu(x)$ and $\Delta(x)$ are smooth, the low-overlap region persists over a broad interval below the bulk topological transition even under very strong disorder. This explains the spectral robustness observed earlier in Figs.~\ref{fig4}(j,k,l): the disorder does not significantly split the mode because the opposite-chirality components are already separated in space, and therefore the disorder-induced matrix element between them remains extremely small. The central conclusion from Figs.~\ref{fig5} and \ref{fig6} is thus that disorder robustness is controlled primarily by the chiral overlap $\Omega$, not directly by the global topological phase boundary.

\section{Wavefunctions of the lowest positive energy and chiral symmetry }\label{sec:5}
\begin{figure}[t!]
    \centering
    \includegraphics[width=1.0\linewidth]{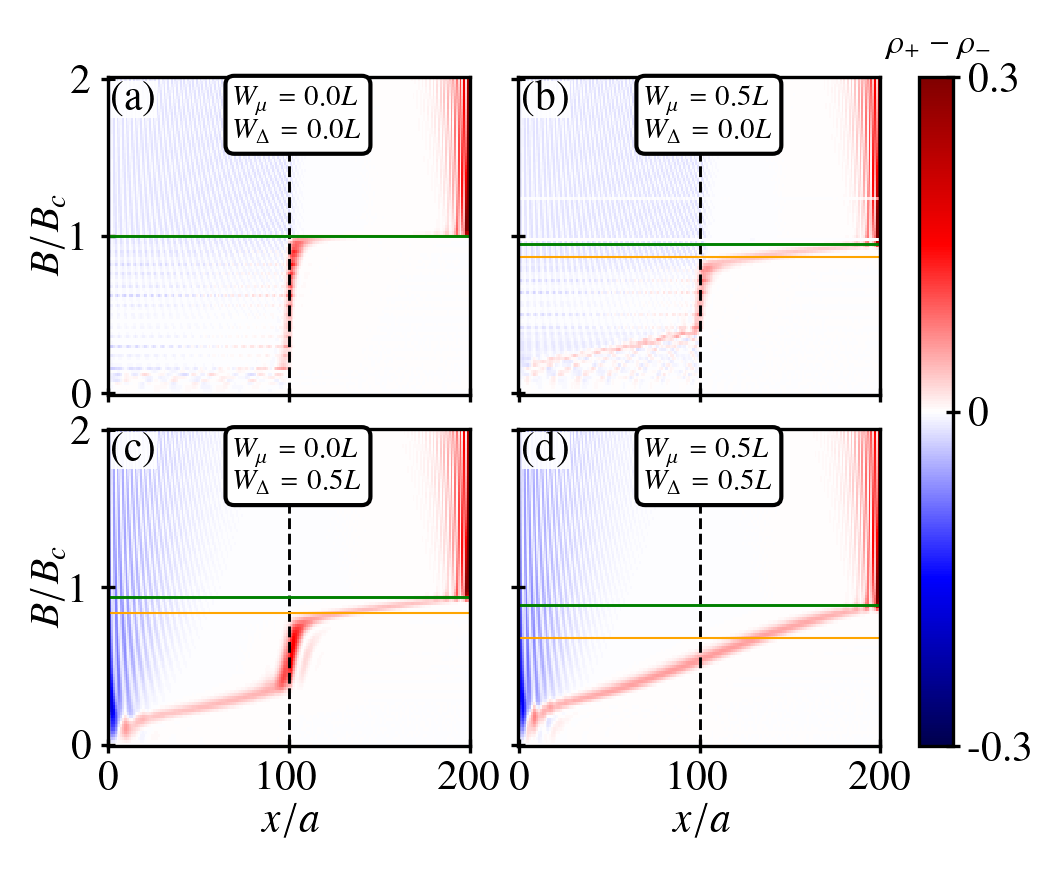}
    \caption{Expectation value of the chiral operator ($\nu(x)$) with respect to the lowest positive energy level as a function of position along the wire and applied Zeeman field with $W_\mu=W_\Delta=0$ (a), $W_\mu = 0.5L, W_\Delta=0$ (b), $W_\mu = 0, W_\Delta=0.5L$ (c), and $W_\mu =W_\Delta=0.5L$ (d).  The green line indicates the bulk gap closing point while the orange line indicates the critical Zeeman field separating disorder-fragile and disorder-robust zero-energy states. }
    \label{fig7}
\end{figure}

\begin{figure}[t!]
    \centering
    \includegraphics[width=1.0\linewidth]{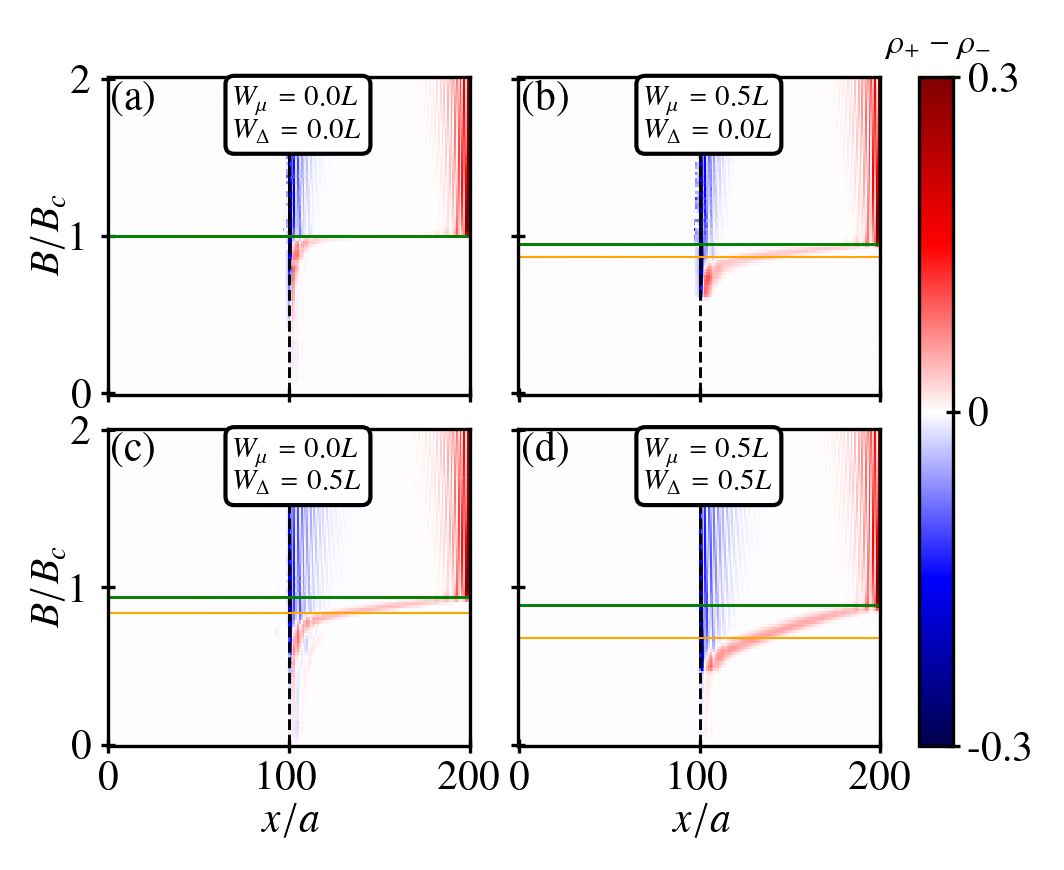}
    \caption{Disorder-averaged expectation value of the chiral operator ($\nu(x)$) with respect to the lowest positive energy level in the presence of strong on-site disorder in the left half of the wire as a function of position along the wire and applied Zeeman field with $W_\mu=W_\Delta=0$ (a), $W_\mu = 0.5L, W_\Delta=0$ (b), $W_\mu = 0, W_\Delta=0.5L$ (c), and $W_\mu =W_\Delta=0.5L$ (d). The green line indicates the bulk gap closing point while the orange line indicates the critical Zeeman field separating disorder-fragile and disorder-robust zero-energy states.}
    \label{fig8}
\end{figure}

While the overlap gives us a hint about the behavior of the wavefunction, it still doesn't explicitly show how the wavefunction rearranges itself in real space as it responds to change of Zeeman field or disorder. To understand the behavior of the wavefunction further, we show the local chiral density $\nu(x)=\rho_+(x) - \rho_-(x)$ as a function of position and applied Zeeman field in the clean (Fig.~\ref{fig7}) and disordered (Fig.~\ref{fig8}) limits.

We first consider the clean system shown in Fig.~\ref{fig7}. In the sharp NS limit, Fig.~\ref{fig7}(a), the trivial low-energy state below the topological transition is concentrated near the NS interface and in the N region. The negative chirality component, identified by negative chiral density $\nu(x)<0$, exclusively occupies the N region and has an oscillatory pattern. Meanwhile, the positive chiral component ($\nu(x)>0$) is predominantly localized at the NS interface. Nevertheless, we note that $\nu(x)\approx0$ inside the N region in the topologically trivial phase owing to the large overlap between the two chiral components of the wavefunction. This remains as a consistent feature throughout the entire trivial regime. Exactly at the bulk topological phase transition --indicated by the solid green line-- the chiral density $\nu(x)$ changes abruptly. The two chiral components suddenly jump to the edges of the system. One chirality is localized near the left edge of the superconducting segment, leaking into the normal segment, while the other appears near the far right end of the superconducting segment. This is the characteristic chirality pattern of a true Majorana pair.  Similar behavior is also observed in Fig.~\ref{fig7}(b) when only the chemical potential profile is smooth. Interestingly, we see that unlike the NS junction with sharp interface case, there is no abrupt change of the chiral density at the topological phase transition. Instead, we see that as the Zeeman field exceeds $B_{c1}$, solid orange line, the interface component (positive chiral component) migrates gradually to the right end of the wire, albeit at rapid rate.

A more pronounced effect appears in Fig.~\ref{fig7}(c), where only the pairing profile is smooth. In this case, the entire wire becomes superconducting but with position-dependent pairing gap. Since the far left end of the wire has the smallest pairing gap, negative chiral component is visibly displaced toward that side for all values of Zeeman field. The positive chiral component initially starts at the same weakly gapped side of the wire but it slowly migrates toward the center of the wire as we increase the Zeeman field. After the positive chiral component reaches the center, it remains at the center until the Zeeman field exceeds $B_{c1}$. We interpret this as a result of the step function profile of the chemical potential. Since the value of the chemical potential jumps as one approaches the right side of the wire, the Zeeman field needs to be strong enough to overcome this jump. This occurs when the Zeeman field is bigger than $B_{c1}$. Only then can the  positive chiral component penetrate the right side of the wire. Above $B_{c1}$, the positive chiral components rapidly moves toward the right end of the wire. It finally reaches the right end at the onset of the topological phase transition.

 Fig.~\ref{fig7}(d), where both   $\mu(x)$ and $\Delta(x)$ are smooth, shares similar characteristics to Fig.~\ref{fig7}(c). Similar to Fig.~\ref{fig7}(c), the negative chirality is bound to the left end of the wire while the positive chirality moves in response to the Zeeman field.  Unlike Fig.~\ref{fig7}(c), both  $\mu(x)$ and $\Delta(x)$ are smooth allowing the positive chiral component to move freely without interruption at the center of the wire. In particular, we see that the location of the positive chiral component is almost linearly proportional to the strength of the Zeeman field in the topologically trivial regime. The positive chiral components enters the right side of the wire near $B=B_{c1}$. Above this value, the two chiral components are well-separated, thereby confirming that $B_{c1}$ is the field at which the trivial low-energy mode changes from an interface-like ABS into a partially separated Majorana-like state.

The same analysis becomes even more illuminating in the disordered system shown in Fig.~\ref{fig8}, especially in the trivial phase. In the NS junction with sharp interface, Fig.~\ref{fig8}(a), the disorder on the left half of the wire effectively cuts the system in half and forces the low-energy states to live inside the remaining right half. Because the trivial state was already strongly overlapping in the clean limit, the positive and negative chiral components become even more overlapping and the chiral density becomes practically zero ($\nu(x)\approx0$). Hence, the near-zero-energy modes are easily lifted away from zero. Similar conclusions apply to the cases where only one profile is smooth, Figs.~\ref{fig8}(b,c). Only in the doubly smooth case, Fig.~\ref{fig8}(d), does the separated chiral structure survive strong disorder over a wide range of Zeeman fields above $B_{c1}$. In that case, the negative chiral component becomes localized at the center of the wire while the positive chiral component is gradually shifted toward the right end of the wire as we increase the Zeeman field.  This persistence of spatial separation under strong perturbations is the real-space origin of the disorder robustness found in Fig.~\ref{fig4}(j,k,l). Disorder strongly reshapes the left part of the wire, but it does not force the two chiral sectors back into the same spatial region, and therefore it cannot efficiently split the zero-energy mode, providing a new robustness mechanism.

\section{Concluding remarks}\label{sec:6}
To conclude, we emphasize that the results presented here have several implications for interpreting zero-energy states in nanowire setups.  First, they show that disorder robustness should not be used as a standalone criterion for global topology.  A state can remain near zero energy under strong disorder because it is a true topological MBS, but it can also remain near zero energy because chiral symmetry and real-space separation suppress the relevant perturbation matrix element.  These mechanisms are related but not identical.  The first is controlled by a bulk topological invariant and a protecting gap.  The second is controlled by the local structure of the wavefunction.
Second, our results provide a symmetry-based interpretation of partially separated ABSs.  Previous studies have emphasized that smooth potentials can generate ABSs whose Majorana components are spatially separated and whose signatures resemble those of topological MBSs \cite{prada2019andreev,marra2022majorana}.  Our contribution is to show that, in a chiral-symmetric setting, the separation of the wavefunction can be expressed as separation of opposite chiralities.  This makes the protection mechanism transparent: chiral-symmetric disorder couples only opposite chiralities, and the corresponding matrix element is bounded by $V_{\max}\Omega$.
Third, the distinction between $B_{c1}$ and $B_{c2}$ suggests a practical way to classify low-energy regimes in inhomogeneous wires.  The field $B_{c2}$ is the conventional bulk topological threshold.  The lower field $B_{c1}$ is not a topological phase transition; it is a wavefunction crossover.  Nevertheless, $B_{c1}$ can mark the onset of experimentally visible robustness.  Therefore, a robust zero-bias feature appearing below the estimated bulk transition should not automatically be dismissed as accidental, nor should it automatically be identified as globally topological.  It may indicate a locally topological, chiral-separated ABS.
Finally, the role of disorder is more nuanced than usual.  Disorder is often treated as a harmful perturbation that destroys Majorana signatures or generates misleading zero-bias anomalies.  Here, because the disorder preserves chiral symmetry, its effect is constrained. It can move and reshape the low-energy wavefunction, effectively shortening locally topological regions or shifting Majorana-like components, but it cannot efficiently split a state whose opposite chiralities have very small overlap.  This observation suggests that controlled disorder or local gates could be used as probes of the internal structure of near-zero-energy states.

In conclusion, we have shown that smooth electrostatic and pairing inhomogeneity can generate a broad regime of disorder-robust, topologically trivial zero-energy states in Rashba nanowires. The key mechanism is the spatial separation of opposite-chirality components of the low-energy wavefunction, which suppresses their hybridization and pins the mode exponentially close to zero energy. The resulting crossover field $B_{c1}$ is distinct from the true topological threshold $B_{c2}$ and marks the onset of a partially separated Majorana-like regime within the trivial phase. Our results provide a unified wavefunction-based picture of robust zero modes in inhomogeneous nanowires and highlight the need to distinguish carefully between global topological order and local Majorana-like structure when interpreting zero-energy signatures in realistic hybrid devices.
\begin{acknowledgments}
    E.A. and Y.T. acknowledge support from JSPS with Grants-in-Aid for Scientific research (KAKENHI Grant No. 25H00613 ). Y.\,T.\, also acknowledges financial support from JSPS with Grants-in-Aid for Scientific Research (KAKENHI Grants No. 23K17668 24K00583 25K07203 25H00609). E.A. also acknowledges financial support from JSPS with Grants-in-Aid for Scientific research (KAKENHI Grant No.	25H01250). J. C. acknowledges financial support from  the Swedish Research Council (Vetenskapsr{\aa}det Grant No. 2021-04121) and from the Olle Engkvist Foundation (Grant No.  243-1026).
    
\end{acknowledgments}

\bibliography{biblio}

@phdthesis{cayao2016hybrid,
  title={Hybrid superconductor-semiconductor nanowire junctions as useful platforms to study {M}ajorana bound states},
  author={Cayao, Jorge},
  year={2016},
  month={May},
  address={Madrid, Spain},
  school={Autonomous University of Madrid (UAM)},
  type         = {PhD thesis},
  note={ [arXiv:1703.07630]}
}

@article{PhysRevB.86.180503,
  title = {Transport spectroscopy of $NS$ nanowire junctions with {M}ajorana fermions},
  author = {Prada, Elsa and San-Jose, Pablo and Aguado, Ram\'on},
  journal = {Phys. Rev. B},
  volume = {86},
  issue = {18},
  pages = {180503},
  numpages = {5},
  year = {2012},
  month = {Nov},
  publisher = {American Physical Society},
  doi = {10.1103/PhysRevB.86.180503},
  url = {https://link.aps.org/doi/10.1103/PhysRevB.86.180503}
}

@article{PhysRevB.98.245407,
  title = {Zero-energy {A}ndreev bound states from quantum dots in proximitized Rashba nanowires},
  author = {Reeg, Christopher and Dmytruk, Olesia and Chevallier, Denis and Loss, Daniel and Klinovaja, Jelena},
  journal = {Phys. Rev. B},
  volume = {98},
  issue = {24},
  pages = {245407},
  numpages = {12},
  year = {2018},
  month = {Dec},
  publisher = {American Physical Society},
  doi = {10.1103/PhysRevB.98.245407},
  url = {https://link.aps.org/doi/10.1103/PhysRevB.98.245407}
}

@article{PhysRevB.86.100503,
  title = {Near-zero-energy end states in topologically trivial spin-orbit coupled superconducting nanowires with a smooth confinement},
  author = {Kells, G. and Meidan, D. and Brouwer, P. W.},
  journal = {Phys. Rev. B},
  volume = {86},
  issue = {10},
  pages = {100503},
  numpages = {5},
  year = {2012},
  month = {Sep},
  publisher = {American Physical Society},
  doi = {10.1103/PhysRevB.86.100503},
  url = {https://link.aps.org/doi/10.1103/PhysRevB.86.100503}
}

@article{PhysRevB.102.245431,
  title = {Pinning of {A}ndreev bound states to zero energy in two-dimensional superconductor- semiconductor Rashba heterostructures},
  author = {Dmytruk, Olesia and Loss, Daniel and Klinovaja, Jelena},
  journal = {Phys. Rev. B},
  volume = {102},
  issue = {24},
  pages = {245431},
  numpages = {7},
  year = {2020},
  month = {Dec},
  publisher = {American Physical Society},
  doi = {10.1103/PhysRevB.102.245431},
  url = {https://link.aps.org/doi/10.1103/PhysRevB.102.245431}
}

@article{Marra_2022,
 doi = {10.1063/5.0102999},
 url = {https://doi.org/10.1063\%2F5.0102999},
 year = 2022,
 month = {dec},
 publisher = {{AIP} Publishing},
 volume = {132},
 number = {23},
 pages = {231101},
 author = {Pasquale Marra},
 title = {Majorana nanowires for topological quantum computation},
 journal = {J. Appl. Phys.}}

@article{tanaka2012symmetry,
  title={Symmetry and topology in superconductors--odd-frequency pairing and edge states--},
  author={Tanaka, Yukio and Sato, Masatoshi and Nagaosa, Naoto},
  journal={J. Phys. Soc. Jpn.},
  volume={81},
  number={1},
  pages={011013},
  year={2012},
  url={https://doi.org/10.1143/JPSJ.81.011013},
  publisher={The Physical Society of Japan}
}

@article{sato2017topological,
  title={Topological superconductors: a review},
  author={Sato, Masatoshi and Ando, Yoichi},
  journal={Rep. Prog. Phys.},
  volume={80},
  number={7},
  pages={076501},
  year={2017},
  url={https://iopscience.iop.org/article/10.1088/1361-6633/aa6ac7/meta},
  publisher={IOP Publishing}
}

@article{Aguadoreview17,
  title={Majorana quasiparticles in condensed matter},
  author={Aguado, Ram{\'o}n},
  journal={Riv. Nuovo Cimento},
   volume = {40},
  issue = {11},
  pages = {523},
  year={2017},
  DOI={10.1393/ncr/i2017-10141-9}
}

@article{lutchyn2018majorana,
  title={Majorana zero modes in superconductor--semiconductor heterostructures},
  author={Lutchyn, Roman M and Bakkers, Erik PAM and Kouwenhoven, Leo P and Krogstrup, Peter and Marcus, Charles M and Oreg, Yuval},
  journal={Nat. Rev. Mater.},
  volume={3},
  number={5},
  pages={52--68},
  year={2018},
  url={https://doi.org/10.1038/s41578-018-0003-1},
  publisher={Nature Publishing Group}
}

@article{flensberg2021engineered,
  title={Engineered platforms for topological superconductivity and {M}ajorana zero modes},
  author={Flensberg, Karsten and von Oppen, Felix and Stern, Ady},
  journal={Nat. Rev. Mater.},
  volume={6},
  number={10},
  pages={944--958},
  year={2021},
  url={https://www.nature.com/articles/s41578-021-00336-6},
  publisher={Nature Publishing Group}
}

@article{frolov2019quest,
  title={Topological superconductivity in hybrid devices},
  author={Frolov, S. M. and Manfra, M. J. and Sau, J. D.},
  journal={Nat. Phys.},
  volume={16},
  number={7},
  pages={718--724},
  year={2020},
  url={https://doi.org/10.1038/s41567-020-0925-6},
  publisher={Nature Publishing Group}
}

@article{zhang2019next,
  title={Next steps of quantum transport in {M}ajorana nanowire devices},
  author={Zhang, Hao and Liu, Dong E and Wimmer, Michael and Kouwenhoven, Leo P},
  journal={Nat. Commun.},
  volume={10},
  number={1},
  pages={5128},
  year={2019},
  publisher={Nature Publishing Group},
  DOI={10.1038/s41467-019-13133-1}
}

@article{prada2019andreev,
  title={From {A}ndreev to {M}ajorana bound states in hybrid superconductor--semiconductor nanowires},
  author={Prada, Elsa and San-Jose, Pablo and de Moor, Michiel WA and Geresdi, Attila and Lee, Eduardo JH and Klinovaja, Jelena and Loss, Daniel and Nyg{\aa}rd, Jesper and Aguado, Ram{\'o}n and Kouwenhoven, Leo P},
  journal={Nat. Rev. Phys.},
  volume={2},
  pages={575--594},
  year={2020},
  url={https://doi.org/10.1038/s42254-020-0228-y},
  publisher={Nature Publishing Group}
}

@article{tanaka2024theory,
 author = {Tanaka, Yukio and Tamura, Shun and Cayao, Jorge},
 title = {Theory of {M}ajorana Zero Modes in Unconventional Superconductors},
 journal = {Prog. Theor. Exp. Phys.},
 volume={2024},
 pages = {08C105},
 year = {2024},
 url = {https://doi.org/10.1093/ptep/ptae065}
}

@article{PhysRevB.104.L020501,
  title = {Distinguishing trivial and topological zero-energy states in long nanowire junctions},
  author = {Cayao, Jorge and Black-Schaffer, Annica M.},
  journal = {Phys. Rev. B},
  volume = {104},
  issue = {2},
  pages = {L020501},
  numpages = {6},
  year = {2021},
  month = {Jul},
  publisher = {American Physical Society},
  doi = {10.1103/PhysRevB.104.L020501},
  url = {https://link.aps.org/doi/10.1103/PhysRevB.104.L020501}
}

@article{cayao2018andreev,
  title={Andreev spectrum and supercurrents in nanowire-based {SNS} junctions containing {M}ajorana bound states},
  author={Cayao, Jorge and Black-Schaffer, Annica M and Prada, Elsa and Aguado, Ram{\'o}n},
  journal={Beilstein J. Nanotechnol.},
  volume={9},
  number={1},
  pages={1339--1357},
  year={2018},
  publisher={Beilstein-Institut},
  url={https://www.beilstein-journals.org/bjnano/articles/9/127},
}

@article{PhysRevLett.123.117001,
  title = {Supercurrent Detection of Topologically Trivial Zero-Energy States in Nanowire Junctions},
  author = {Awoga, Oladunjoye A. and Cayao, Jorge and Black-Schaffer, Annica M.},
  journal = {Phys. Rev. Lett.},
  volume = {123},
  issue = {11},
  pages = {117001},
  numpages = {7},
  year = {2019},
  month = {Sep},
  publisher = {American Physical Society},
  doi = {10.1103/PhysRevLett.123.117001},
  url = {https://link.aps.org/doi/10.1103/PhysRevLett.123.117001}
}

@article{PhysRevB.110.165404,
  title = {Identifying trivial and Majorana zero-energy modes using the Majorana polarization},
  author = {Awoga, Oladunjoye A. and Cayao, Jorge},
  journal = {Phys. Rev. B},
  volume = {110},
  issue = {16},
  pages = {165404},
  numpages = {11},
  year = {2024},
  month = {Oct},
  publisher = {American Physical Society},
  doi = {10.1103/PhysRevB.110.165404},
  url = {https://link.aps.org/doi/10.1103/PhysRevB.110.165404}
}

@article{JorgeEPs,
	author = {Pablo San-Jos\'{e} and Jorge Cayao and Elsa Prada and Ram{\'o}n Aguado},
	journal = {Sci. Rep.},
	title = {Majorana bound states from exceptional points in non-topological superconductors},
		pages = {21427},
	volume = {6},
	url={http://dx.doi.org/10.1038/srep21427},
	year = {2016}}

@article{Bagrets:PRL12,
  title = {Class {$D$} Spectral Peak in {M}ajorana Quantum Wires},
  author = {Bagrets, Dmitry and Altland, Alexander},
  journal = {Phys. Rev. Lett.},
  volume = {109},
  issue = {22},
  pages = {227005},
  numpages = {5},
  year = {2012},
  month = {Nov},
  publisher = {American Physical Society},
  url = {https://link.aps.org/doi/10.1103/PhysRevLett.109.227005}
}

@article{PhysRevB.104.134507,
  title = {Confinement-induced zero-bias peaks in conventional superconductor hybrids},
  author = {Cayao, Jorge and Burset, Pablo},
  journal = {Phys. Rev. B},
  volume = {104},
  issue = {13},
  pages = {134507},
  numpages = {8},
  year = {2021},
  month = {Oct},
  publisher = {American Physical Society},
  doi = {10.1103/PhysRevB.104.134507},
  url = {https://link.aps.org/doi/10.1103/PhysRevB.104.134507}
}

@article{Pikulin2012A,
	doi = {10.1088/1367-2630/14/12/125011},
	url = {https://doi.org/10.1088/1367-2630/14/12/125011},
	year = 2012,
	publisher = {{IOP} Publishing},
	volume = {14},
	number = {12},
	pages = {125011},
	author = {D I Pikulin and J P Dahlhaus and M Wimmer and H Schomerus and C W J Beenakker},
	title = {A zero-voltage conductance peak from weak antilocalization in a {M}ajorana nanowire},
	journal = {New J. Phys.}
}

@article{PhysRevB.91.024514,
  title = {SNS junctions in nanowires with spin-orbit coupling: Role of confinement and helicity on the subgap spectrum},
  author = {Cayao, Jorge and Prada, Elsa and San-Jose, Pablo and Aguado, Ram\'on},
  journal = {Phys. Rev. B},
  volume = {91},
  issue = {2},
  pages = {024514},
  numpages = {15},
  year = {2015},
  month = {Jan},
  publisher = {American Physical Society},
  doi = {10.1103/PhysRevB.91.024514},
  url = {https://link.aps.org/doi/10.1103/PhysRevB.91.024514}
}

@article{sarma2015majorana,
  title={Majorana zero modes and topological quantum computation},
  author={Sarma, Sankar Das and Freedman, Michael and Nayak, Chetan},
  journal={npj Quantum Inf.},
  volume={1},
  number={1},
  pages={15001},
  year={2015},
  url={https://doi.org/10.1038/npjqi.2015.1},
  publisher={Nature Publishing Group}
}

@article{DasSarma2021Disorder,
  title = {Disorder-induced zero-bias peaks in {M}ajorana nanowires},
  author = {Das Sarma, Sankar and Pan, Haining},
  journal = {Phys. Rev. B},
  volume = {103},
  issue = {19},
  pages = {195158},
  numpages = {12},
  year = {2021},
  month = {May},
  publisher = {American Physical Society},
  doi = {10.1103/PhysRevB.103.195158},
  url = {https://link.aps.org/doi/10.1103/PhysRevB.103.195158}
}

@article{PhysRevB.105.144509,
  title = {Robust topological superconductivity in weakly coupled nanowire-superconductor hybrid structures},
  author = {Awoga, Oladunjoye A. and Cayao, Jorge and Black-Schaffer, Annica M.},
  journal = {Phys. Rev. B},
  volume = {105},
  issue = {14},
  pages = {144509},
  numpages = {10},
  year = {2022},
  publisher = {American Physical Society},
  doi = {10.1103/PhysRevB.105.144509},
  url = {https://link.aps.org/doi/10.1103/PhysRevB.105.144509}
}

@article{PhysRevB.107.184519,
  title = {Mitigating disorder-induced zero-energy states in weakly coupled superconductor-semiconductor hybrid systems},
  author = {Awoga, Oladunjoye A. and Leijnse, Martin and Black-Schaffer, Annica M. and Cayao, Jorge},
  journal = {Phys. Rev. B},
  volume = {107},
  issue = {18},
  pages = {184519},
  numpages = {10},
  year = {2023},
  month = {May},
  publisher = {American Physical Society},
  doi = {10.1103/PhysRevB.107.184519},
  url = {https://link.aps.org/doi/10.1103/PhysRevB.107.184519}
}

@article{baldo2023zero,
   title={Zero-frequency supercurrent susceptibility signatures of trivial and topological zero-energy states in nanowire junctions},
   volume={36},
   url={http://dx.doi.org/10.1088/1361-6668/acb670},
   DOI={10.1088/1361-6668/acb670},
   number={3},
   journal={Supercond. Sci. Technol.},
   author={Baldo, Lucas and Dias Da Silva, Luis G G V and Black-Schaffer, Annica M and Cayao, Jorge},
   year={2023},
  pages={034003} }

@article{marra2022majorana,
  title={Majorana/Andreev crossover and the fate of the topological phase transition in inhomogeneous nanowires},
  author={Marra, Pasquale and Nigro, Angela},
  journal={J. Phys.: Condens. Matter},
  volume={34},
  number={12},
  pages={124001},
  year={2022},
  url={https://iopscience.iop.org/article/10.1088/1361-648X/ac44d2},
  publisher={IOP Publishing}
}

@article{PhysRevB.87.104513,
  title = {Majorana fermions and odd-frequency {C}ooper pairs in a normal-metal nanowire proximity-coupled to a topological superconductor},
  author = {Asano, Yasuhiro and Tanaka, Yukio},
  journal = {Phys. Rev. B},
  volume = {87},
  issue = {10},
  pages = {104513},
  numpages = {10},
  year = {2013},
  month = {Mar},
  publisher = {American Physical Society},
  doi = {10.1103/PhysRevB.87.104513},
  url = {https://link.aps.org/doi/10.1103/PhysRevB.87.104513}
}

@article{Proximityp,
  title = {Anomalous charge transport in triplet superconductor junctions},
  author = {Tanaka, Y. and Kashiwaya, S.},
  journal = {Phys. Rev. B},
  volume = {70},
  issue = {1},
  pages = {012507},
  numpages = {4},
  year = {2004},
  month = {Jul},
  publisher = {American Physical Society},
  doi = {10.1103/PhysRevB.70.012507},
  url = {https://link.aps.org/doi/10.1103/PhysRevB.70.012507}
}

@article{PhysRevB.71.094513,
  title = {Theory of enhanced proximity effect by midgap Andreev resonant state in diffusive normal-metal/triplet superconductor junctions},
  author = {Tanaka, Y. and Kashiwaya, S. and Yokoyama, T.},
  journal = {Phys. Rev. B},
  volume = {71},
  issue = {9},
  pages = {094513},
  numpages = {16},
  year = {2005},
  month = {Mar},
  publisher = {American Physical Society},
  doi = {10.1103/PhysRevB.71.094513},
  url = {https://link.aps.org/doi/10.1103/PhysRevB.71.094513}
}

@article{odd1,
  title = {Theory of the Proximity Effect in Junctions with Unconventional Superconductors},
  author = {Tanaka, Y. and Golubov, A. A.},
  journal = {Phys. Rev. Lett.},
  volume = {98},
  issue = {3},
  pages = {037003},
  numpages = {4},
  year = {2007},
  month = {Jan},
  publisher = {American Physical Society},
  doi = {10.1103/PhysRevLett.98.037003},
  url = {https://link.aps.org/doi/10.1103/PhysRevLett.98.037003}
}

@article{odd3,
  title = {Anomalous Josephson Effect between Even- and Odd-Frequency Superconductors},
  author = {Tanaka, Yukio and Golubov, Alexander A. and Kashiwaya, Satoshi and Ueda, Masahito},
  journal = {Phys. Rev. Lett.},
  volume = {99},
  issue = {3},
  pages = {037005},
  numpages = {4},
  year = {2007},
  month = {Jul},
  publisher = {American Physical Society},
  doi = {10.1103/PhysRevLett.99.037005},
  url = {https://link.aps.org/doi/10.1103/PhysRevLett.99.037005}
}

@article{Spectralbulk,
  title = {Odd-frequency pairs in chiral symmetric systems: Spectral bulk-boundary correspondence and topological criticality},
  author = {Tamura, Shun and Hoshino, Shintaro and Tanaka, Yukio},
  journal = {Phys. Rev. B},
  volume = {99},
  issue = {18},
  pages = {184512},
  numpages = {17},
  year = {2019},
  month = {May},
  publisher = {American Physical Society},
  doi = {10.1103/PhysRevB.99.184512},
  url = {https://link.aps.org/doi/10.1103/PhysRevB.99.184512}
}

@article{Oreg,
  title = {Helical Liquids and Majorana Bound States in Quantum Wires},
  author = {Oreg, Yuval and Refael, Gil and von Oppen, Felix},
  journal = {Phys. Rev. Lett.},
  volume = {105},
  issue = {17},
  pages = {177002},
  numpages = {4},
  year = {2010},
  month = {Oct},
  publisher = {American Physical Society},
  doi = {10.1103/PhysRevLett.105.177002},
  url = {https://link.aps.org/doi/10.1103/PhysRevLett.105.177002}
}

@article{Lutchyn,
  title = {Majorana Fermions and a Topological Phase Transition in Semiconductor-Superconductor Heterostructures},
  author = {Lutchyn, Roman M. and Sau, Jay D. and Das Sarma, S.},
  journal = {Phys. Rev. Lett.},
  volume = {105},
  issue = {7},
  pages = {077001},
  numpages = {4},
  year = {2010},
  month = {Aug},
  publisher = {American Physical Society},
  doi = {10.1103/PhysRevLett.105.077001},
  url = {https://link.aps.org/doi/10.1103/PhysRevLett.105.077001}
}

@article{Alicea_2012,
doi = {10.1088/0034-4885/75/7/076501},
url = {https://dx.doi.org/10.1088/0034-4885/75/7/076501},
year = {2012},
month = {jun},
publisher = {IOP Publishing},
volume = {75},
number = {7},
pages = {076501},
author = {Jason Alicea},
title = {New directions in the pursuit of Majorana fermions in solid state systems},
journal = {Reports on Progress in Physics}
}

@article{Beenakker_2013,
author = {Beenakker, C.W.J.},
title = {Search for Majorana Fermions in Superconductors},
journal = {Annual Review of Condensed Matter Physics},
volume = {4},
number = {1},
pages = {113-136},
year = {2013},
doi = {10.1146/annurev-conmatphys-030212-184337},
URL = {https://doi.org/10.1146/annurev-conmatphys-030212-184337}
}

@article{Stanescu_2013,
doi = {10.1088/0953-8984/25/23/233201},
url = {https://dx.doi.org/10.1088/0953-8984/25/23/233201},
year = {2013},
month = {may},
publisher = {IOP Publishing},
volume = {25},
number = {23},
pages = {233201},
author = {T D Stanescu and S Tewari},
title = {Majorana fermions in semiconductor nanowires: fundamentals, modeling, and experiment},
journal = {Journal of Physics: Condensed Matter}
}

@article{PhysRevB.72.140503,
  title = {Anomalous features of the proximity effect in triplet superconductors},
  author = {Tanaka, Y. and Asano, Y. and Golubov, A. A. and Kashiwaya, S.},
  journal = {Phys. Rev. B},
  volume = {72},
  issue = {14},
  pages = {140503},
  numpages = {4},
  year = {2005},
  month = {Oct},
  publisher = {American Physical Society},
  doi = {10.1103/PhysRevB.72.140503},
  url = {https://link.aps.org/doi/10.1103/PhysRevB.72.140503}
}

@article{PhysRevB.91.174511,
  title = {Anomalous proximity effect and theoretical design for its realization},
  author = {Ikegaya, Satoshi and Asano, Yasuhiro and Tanaka, Yukio},
  journal = {Phys. Rev. B},
  volume = {91},
  issue = {17},
  pages = {174511},
  numpages = {6},
  year = {2015},
  month = {May},
  publisher = {American Physical Society},
  doi = {10.1103/PhysRevB.91.174511},
  url = {https://link.aps.org/doi/10.1103/PhysRevB.91.174511}
}

@article{PhysRevB.94.054512,
  title = {Quantization of conductance minimum and index theorem},
  author = {Ikegaya, Satoshi and Suzuki, Shu-Ichiro and Tanaka, Yukio and Asano, Yasuhiro},
  journal = {Phys. Rev. B},
  volume = {94},
  issue = {5},
  pages = {054512},
  numpages = {6},
  year = {2016},
  month = {Aug},
  publisher = {American Physical Society},
  doi = {10.1103/PhysRevB.94.054512},
  url = {https://link.aps.org/doi/10.1103/PhysRevB.94.054512}
}

@article{PhysRevB.95.214503,
  title = {Stability of flat zero-energy states at the dirty surface of a nodal superconductor},
  author = {Ikegaya, Satoshi and Asano, Yasuhiro},
  journal = {Phys. Rev. B},
  volume = {95},
  issue = {21},
  pages = {214503},
  numpages = {11},
  year = {2017},
  month = {Jun},
  publisher = {American Physical Society},
  doi = {10.1103/PhysRevB.95.214503},
  url = {https://link.aps.org/doi/10.1103/PhysRevB.95.214503}
}

@article{PhysRevB.97.174501,
  title = {Symmetry conditions of a nodal superconductor for generating robust flat-band Andreev bound states at its dirty surface},
  author = {Ikegaya, Satoshi and Kobayashi, Shingo and Asano, Yasuhiro},
  journal = {Phys. Rev. B},
  volume = {97},
  issue = {17},
  pages = {174501},
  numpages = {9},
  year = {2018},
  month = {May},
  publisher = {American Physical Society},
  doi = {10.1103/PhysRevB.97.174501},
  url = {https://link.aps.org/doi/10.1103/PhysRevB.97.174501}
}

@article{PhysRevB.102.140505,
  title = {Anomalous proximity effect of planar topological Josephson junctions},
  author = {Ikegaya, S. and Tamura, S. and Manske, D. and Tanaka, Y.},
  journal = {Phys. Rev. B},
  volume = {102},
  issue = {14},
  pages = {140505},
  numpages = {5},
  year = {2020},
  month = {Oct},
  publisher = {American Physical Society},
  doi = {10.1103/PhysRevB.102.140505},
  url = {https://link.aps.org/doi/10.1103/PhysRevB.102.140505}
}

@article{ahmed2025odd,
  title={Odd-frequency pairing due to Majorana and trivial Andreev bound states},
  author={Ahmed, Eslam and Tamura, Shun and Tanaka, Yukio and Cayao, Jorge},
  journal={Physical Review B},
  volume={111},
  number={22},
  pages={224508},
  year={2025},
  publisher={APS}
}

@article{Mizushima2023,
  title = {Odd-frequency pairs and anomalous proximity effect in nematic and chiral states of superconducting topological insulators},
  author = {Mizushima, Takeshi and Tamura, Shun and Yada, Keiji and Tanaka, Yukio},
  journal = {Phys. Rev. B},
  volume = {107},
  issue = {6},
  pages = {064504},
  numpages = {27},
  year = {2023},
  month = {Feb},
  publisher = {American Physical Society},
  doi = {10.1103/PhysRevB.107.064504},
  url = {https://link.aps.org/doi/10.1103/PhysRevB.107.064504}
}

@article{NagaeFlatband2025,
  title = {Flat-band zero-energy states and anomalous proximity effects in $p$-wave magnet--superconductor hybrid systems},
  author = {Nagae, Yutaro and Katayama, Leo and Ikegaya, Satoshi},
  journal = {Phys. Rev. B},
  volume = {111},
  issue = {17},
  pages = {174519},
  numpages = {11},
  year = {2025},
  month = {May},
  publisher = {American Physical Society},
  doi = {10.1103/PhysRevB.111.174519},
  url = {https://link.aps.org/doi/10.1103/PhysRevB.111.174519}
}

@article{Kokkeler2022,
  title = {Theory of proximity effect in $s+p$-wave superconductor junctions},
  author = {Tanaka, Yukio and Kokkeler, Tim and Golubov, Alexander},
  journal = {Phys. Rev. B},
  volume = {105},
  issue = {21},
  pages = {214512},
  numpages = {21},
  year = {2022},
  month = {Jun},
  publisher = {American Physical Society},
  doi = {10.1103/PhysRevB.105.214512},
  url = {https://link.aps.org/doi/10.1103/PhysRevB.105.214512}
}

@article{Kitaev_2001,
doi = {10.1070/1063-7869/44/10S/S29},
url = {https://dx.doi.org/10.1070/1063-7869/44/10S/S29},
year = {2001},
month = {oct},
publisher = {},
volume = {44},
number = {10S},
pages = {131},
author = {A Yu Kitaev},
title = {Unpaired Majorana fermions in quantum
wires},
journal = {Physics-Uspekhi}
}

@article{ahmed2025anomalous,
  title={Anomalous proximity effect under Andreev and majorana bound states},
  author={Ahmed, Eslam and Tanaka, Yukio and Cayao, Jorge},
  journal={Journal of Superconductivity and Novel Magnetism},
  volume={38},
  number={5},
  pages={220},
  year={2025},
  month={10},
  publisher={Springer},
  doi={10.1007/s10948-025-07057-9}
}

@article{PhysRevB.98.235406,
  title = {Quantifying wave-function overlaps in inhomogeneous Majorana nanowires},
  author = {Pe\~naranda, Fernando and Aguado, Ram\'on and San-Jose, Pablo and Prada, Elsa},
  journal = {Phys. Rev. B},
  volume = {98},
  issue = {23},
  pages = {235406},
  numpages = {14},
  year = {2018},
  month = {Dec},
  publisher = {American Physical Society},
  doi = {10.1103/PhysRevB.98.235406},
  url = {https://link.aps.org/doi/10.1103/PhysRevB.98.235406}
}

@article{PhysRevB.105.205122,
  title = {Partially separated Majorana modes in a disordered medium},
  author = {Zeng, Chuanchang and Sharma, Gargee and Tewari, Sumanta and Stanescu, Tudor},
  journal = {Phys. Rev. B},
  volume = {105},
  issue = {20},
  pages = {205122},
  numpages = {12},
  year = {2022},
  month = {May},
  publisher = {American Physical Society},
  doi = {10.1103/PhysRevB.105.205122},
  url = {https://link.aps.org/doi/10.1103/PhysRevB.105.205122}
}

@article{PhysRevB.84.144526,
  title = {Topological superconducting phases in disordered quantum wires with strong spin-orbit coupling},
  author = {Brouwer, Piet W. and Duckheim, Mathias and Romito, Alessandro and von Oppen, Felix},
  journal = {Phys. Rev. B},
  volume = {84},
  issue = {14},
  pages = {144526},
  numpages = {6},
  year = {2011},
  month = {Oct},
  publisher = {American Physical Society},
  doi = {10.1103/PhysRevB.84.144526},
  url = {https://link.aps.org/doi/10.1103/PhysRevB.84.144526}
}

@article{PhysRevLett.106.057001,
  title = {Quantized Conductance at the Majorana Phase Transition in a Disordered Superconducting Wire},
  author = {Akhmerov, A. R. and Dahlhaus, J. P. and Hassler, F. and Wimmer, M. and Beenakker, C. W. J.},
  journal = {Phys. Rev. Lett.},
  volume = {106},
  issue = {5},
  pages = {057001},
  numpages = {4},
  year = {2011},
  month = {Jan},
  publisher = {American Physical Society},
  doi = {10.1103/PhysRevLett.106.057001},
  url = {https://link.aps.org/doi/10.1103/PhysRevLett.106.057001}
}

@article{PhysRevB.83.184520,
  title = {Engineering a $p+\mathit{ip}$ superconductor: Comparison of topological insulator and Rashba spin-orbit-coupled materials},
  author = {Potter, Andrew C. and Lee, Patrick A.},
  journal = {Phys. Rev. B},
  volume = {83},
  issue = {18},
  pages = {184520},
  numpages = {11},
  year = {2011},
  month = {May},
  publisher = {American Physical Society},
  doi = {10.1103/PhysRevB.83.184520},
  url = {https://link.aps.org/doi/10.1103/PhysRevB.83.184520}
}

@article{PhysRevB.85.140513,
  title = {Momentum relaxation in a semiconductor proximity-coupled to a disordered $s$-wave superconductor: Effect of scattering on topological superconductivity},
  author = {Lutchyn, Roman M. and Stanescu, Tudor D. and Das Sarma, S.},
  journal = {Phys. Rev. B},
  volume = {85},
  issue = {14},
  pages = {140513(R)},
  numpages = {5},
  year = {2012},
  month = {Apr},
  publisher = {American Physical Society},
  doi = {10.1103/PhysRevB.85.140513},
  url = {https://link.aps.org/doi/10.1103/PhysRevB.85.140513}
}

@article{PhysRevLett.109.227006,
  title = {Enhanced Zero-Bias Majorana Peak in the Differential Tunneling Conductance of Disordered Multisubband Quantum-Wire/Superconductor Junctions},
  author = {Pientka, Falko and Kells, Graham and Romito, Alessandro and Brouwer, Piet W. and von Oppen, Felix},
  journal = {Phys. Rev. Lett.},
  volume = {109},
  issue = {22},
  pages = {227006},
  numpages = {5},
  year = {2012},
  month = {Nov},
  publisher = {American Physical Society},
  doi = {10.1103/PhysRevLett.109.227006},
  url = {https://link.aps.org/doi/10.1103/PhysRevLett.109.227006}
}

@article{PhysRevB.88.064506,
  title = {Density of states of disordered topological superconductor-semiconductor hybrid nanowires},
  author = {Sau, Jay D. and Das Sarma, S.},
  journal = {Phys. Rev. B},
  volume = {88},
  issue = {6},
  pages = {064506},
  numpages = {7},
  year = {2013},
  month = {Aug},
  publisher = {American Physical Society},
  doi = {10.1103/PhysRevB.88.064506},
  url = {https://link.aps.org/doi/10.1103/PhysRevB.88.064506}
}

@article{PhysRevB.94.140505,
  title = {Proximity effect and Majorana bound states in clean semiconductor nanowires coupled to disordered superconductors},
  author = {Cole, William S. and Sau, Jay D. and Das Sarma, S.},
  journal = {Phys. Rev. B},
  volume = {94},
  issue = {14},
  pages = {140505(R)},
  numpages = {5},
  year = {2016},
  month = {Oct},
  publisher = {American Physical Society},
  doi = {10.1103/PhysRevB.94.140505},
  url = {https://link.aps.org/doi/10.1103/PhysRevB.94.140505}
}

@misc{prodanov2026interfaces,
      title={Majorana fermions at self-generated interfaces}, 
      author={Nikola Prodanov and Sergio Ciuchi and Sergio Caprara},
      year={2026},
      eprint={2606.10812},
      archivePrefix={arXiv},
      primaryClass={cond-mat.supr-con},
      url={https://arxiv.org/abs/2606.10812}, 
}

@article{sato2011topology,
  title = {Topology of Andreev bound states with flat dispersion},
  author = {Sato, Masatoshi and Tanaka, Yukio and Yada, Keiji and Yokoyama, Takehito},
  journal = {Phys. Rev. B},
  volume = {83},
  issue = {22},
  pages = {224511},
  numpages = {22},
  year = {2011},
  month = {Jun},
  publisher = {American Physical Society},
  doi = {10.1103/PhysRevB.83.224511},
  url = {https://link.aps.org/doi/10.1103/PhysRevB.83.224511}
}

@article{PhysRevB.100.174512,
  title = {Chirality polarizations and spectral bulk-boundary correspondence},
  author = {Daido, Akito and Yanase, Youichi},
  journal = {Phys. Rev. B},
  volume = {100},
  issue = {17},
  pages = {174512},
  numpages = {14},
  year = {2019},
  month = {Nov},
  publisher = {American Physical Society},
  doi = {10.1103/PhysRevB.100.174512},
  url = {https://link.aps.org/doi/10.1103/PhysRevB.100.174512}
}

@article{PhysRevB.104.165125,
  title = {Generalization of spectral bulk-boundary correspondence},
  author = {Tamura, Shun and Hoshino, Shintaro and Tanaka, Yukio},
  journal = {Phys. Rev. B},
  volume = {104},
  issue = {16},
  pages = {165125},
  numpages = {15},
  year = {2021},
  month = {Oct},
  publisher = {American Physical Society},
  doi = {10.1103/PhysRevB.104.165125},
  url = {https://link.aps.org/doi/10.1103/PhysRevB.104.165125}
}

@article{Mizushima_review2016,
author = {Mizushima ,Takeshi and Tsutsumi ,Yasumasa and Kawakami ,Takuto and Sato ,Masatoshi and Ichioka ,Masanori and Machida ,Kazushige},
title = {Symmetry-Protected Topological Superfluids and Superconductors —From the Basics to 3He—},
journal = {Journal of the Physical Society of Japan},
volume = {85},
number = {2},
pages = {022001},
year = {2016},
doi = {10.7566/JPSJ.85.022001},
URL = {https://doi.org/10.7566/JPSJ.85.022001},
eprint = {https://doi.org/10.7566/JPSJ.85.022001}
}

@article{Klinovaja_review2021,
    author = {Laubscher, Katharina and Klinovaja, Jelena},
    title = {Majorana bound states in semiconducting nanostructures},
    journal = {Journal of Applied Physics},
    volume = {130},
    number = {8},
    pages = {081101},
    year = {2021},
    month = {08},
    issn = {0021-8979},
    doi = {10.1063/5.0055997},
    url = {https://doi.org/10.1063/5.0055997},
}

@article{pita2025novel,
  title={Novel qubits in hybrid semiconductor-superconductor nanostructures},
  author={Pita-Vidal, Marta and Souto, Rub{\'e}n Seoane and Goswami, Srijit and Andersen, Christian Kraglund and Katsaros, Georgios and Shabani, Javad and Aguado, Ram{\'o}n},
  journal={arXiv preprint arXiv:2512.23336},
  year={2025}
}

@article{PhysRevB.90.064513,
  title = {Anomalous surface states at interfaces in $p$-wave superconductors},
  author = {Bakurskiy, S. V. and Golubov, A. A. and Kupriyanov, M. Yu. and Yada, K. and Tanaka, Y.},
  journal = {Phys. Rev. B},
  volume = {90},
  issue = {6},
  pages = {064513},
  numpages = {10},
  year = {2014},
  month = {Aug},
  publisher = {American Physical Society},
  doi = {10.1103/PhysRevB.90.064513},
  url = {https://link.aps.org/doi/10.1103/PhysRevB.90.064513}
}

@article{PhysRevLett.108.096802,
  title = {Spin and Majorana Polarization in Topological Superconducting Wires},
  author = {Sticlet, Doru and Bena, Cristina and Simon, Pascal},
  journal = {Phys. Rev. Lett.},
  volume = {108},
  issue = {9},
  pages = {096802},
  numpages = {5},
  year = {2012},
  month = {Mar},
  publisher = {American Physical Society},
  doi = {10.1103/PhysRevLett.108.096802},
  url = {https://link.aps.org/doi/10.1103/PhysRevLett.108.096802}
}

@article{PhysRevB.92.115115,
  title = {Visualizing Majorana bound states in one and two dimensions using the generalized Majorana polarization},
  author = {Sedlmayr, N. and Bena, C.},
  journal = {Phys. Rev. B},
  volume = {92},
  issue = {11},
  pages = {115115},
  numpages = {7},
  year = {2015},
  month = {Sep},
  publisher = {American Physical Society},
  doi = {10.1103/PhysRevB.92.115115},
  url = {https://link.aps.org/doi/10.1103/PhysRevB.92.115115}
}

@article{PhysRevB.93.155425,
  title = {Majorana bound states in open quasi-one-dimensional and two-dimensional systems with transverse Rashba coupling},
  author = {Sedlmayr, N. and Aguiar-Hualde, J. M. and Bena, C.},
  journal = {Phys. Rev. B},
  volume = {93},
  issue = {15},
  pages = {155425},
  numpages = {13},
  year = {2016},
  month = {Apr},
  publisher = {American Physical Society},
  doi = {10.1103/PhysRevB.93.155425},
  url = {https://link.aps.org/doi/10.1103/PhysRevB.93.155425}
}

@article{bena_2017,
     author = {Cristina Bena},
     title = {Testing the formation of {Majorana} states using {Majorana} polarization},
     journal = {Comptes Rendus. Physique},
     pages = {349--357},
     year = {2017},
     publisher = {Elsevier},
     volume = {18},
     number = {5-6},
     doi = {10.1016/j.crhy.2017.09.005},
}

@article{Kaladzhyan_2017,
   title={Majorana fermions in finite-size strips with in-plane magnetic fields},
   volume={90},
   ISSN={1434-6036},
   url={http://dx.doi.org/10.1140/epjb/e2017-80103-y},
   DOI={10.1140/epjb/e2017-80103-y},
   number={11},
   journal={The European Physical Journal B},
   publisher={Springer Science and Business Media LLC},
   author={Kaladzhyan, Vardan and Despres, Julien and Mandal, Ipsita and Bena, Cristina},
   year={2017},
   month=Nov }
\end{document}